\documentclass[lettersize,journal]{IEEEtran}
\usepackage{amsmath,amsfonts}
\usepackage{amssymb}
\usepackage{algorithm}
\usepackage{algorithmic}
\usepackage{array}
\usepackage[caption=false,font=normalsize,labelfont=sf,textfont=sf]{subfig}
\usepackage{textcomp}
\usepackage{stfloats}
\usepackage{color}
\usepackage{url}
\usepackage{verbatim}
\usepackage{cite}
\usepackage{mathrsfs}
\usepackage{bbding}
\usepackage{fontawesome}
\usepackage{subcaption}
\usepackage{pifont}
\usepackage{graphicx}
\usepackage{subfig}

\newtheorem{remark}{\textbf{Remark}} 

\begin{document}

\title{\Huge Tri-Hybrid Holographic Beamforming for Integrated Sensing and Communication}

\author{
\IEEEauthorblockN
{
Shupei Zhang, \IEEEmembership{Graduate Student Member, IEEE}, Shuhao Zeng, \IEEEmembership{Member, IEEE},\\
Boya Di, \IEEEmembership{Senior Member, IEEE}, and  Lingyang Song, \IEEEmembership{Fellow, IEEE}
}
\thanks{Shupei Zhang, Boya Di and Lingyang Song are with State Key Laboratory of Photonics and Communications, School of Electronics, Peking University, Beijing, China (emails: zhangshupei@pku.edu.cn; diboya@pku.edu.cn; lingyang.song@pku.edu.cn).

Shuhao Zeng is with School of Electronic and Computer Engineering, Peking University Shenzhen Graduate School, Shenzhen 518055, China (email: shuhao.zeng96@gmail.com).

Lingyang Song is also with School of Electronic and Computer Engineering, Peking University Shenzhen Graduate School, Shenzhen 518055, China.
}
}
\maketitle

\begin{abstract}
Integrated sensing and communication~(ISAC) can perform both communication and sensing tasks using the same frequency band and hardware, making it a key technology for 6G. As a low-cost implementation for large-scale transceiver antenna arrays, reconfigurable holographic surfaces~(RHSs) can be integrated into ISAC systems to realize the \emph{holographic ISAC} paradigm, where enlarged radiation apertures achieve significant beamforming gains.
In this paper, we investigate the \emph{tri-hybrid} holographic ISAC framework, where the beamformer comprises digital, analog, and RHS-based electromagnetic~(EM) layers.
The analog layer employs a small number of phase shifters~(PSs) to provide subarray-level phase control for the amplitude-modulated RHSs.
Tri-hybrid beamforming provides a pathway for low-cost large-scale holographic ISAC.
However, compared to conventional ISAC systems, it is challenging to achieve joint subarray-level phase control via PSs and element-level radiation amplitude manipulation via RHSs for holographic ISAC.
To address this, we present a tri-hybrid holographic ISAC scheme that minimizes sensing waveform error while satisfying the minimum user rate requirement.
A joint optimization approach for PS phases and RHS amplitude responses is designed to address inter-layer coupling and distinct feasible regions.
Theoretical analyses reveal that the optimized amplitude responses cluster near boundary values, i.e., 1-bit amplitude control, to reduce hardware and algorithmic complexity.
Simulation results show that the proposed scheme achieves a controllable performance trade-off between communication and sensing tasks.
Measured RHS beam gain validates the enhancement of holographic beamforming through subarray-level phase shifting.
Moreover, as the number of RHS elements increases, the proposed approach exceeds the performance of conventional hybrid beamforming while significantly reducing the number of PSs.
\end{abstract}


\section{Introduction}

With the emergence of new applications such as autonomous driving and digital twins, future 6G networks are expected to possess not only enhanced communication capabilities but also robust sensing functionalities~\cite{6G2}. Integrated sensing and communication~(ISAC) is recognized as a pivotal technology for 6G, enabling wireless systems to deliver data transmission while simultaneously performing sensing~\cite{6G_ISAC1}. By utilizing unified waveforms and signals to perform both tasks within the same frequency band, ISAC improves spectrum utilization~\cite{6G_ISAC2}. Moreover, it eliminates the need for deploying separate hardware infrastructures for communication and sensing respectively, as a single ISAC base station~(BS) can fulfill both roles, thereby reducing hardware costs~\cite{6G_ISAC3}.

Owing to the significantly enhanced spatial resolution and spectral efficiency, large-scale arrays provide an effective pathway to enable simultaneous transmission and sensing~\cite{ELA2}. Specifically, the large number of antenna elements improves the beamforming capability of the BS, creating opportunities for integrating sensing waveforms under constrained communication resources~\cite{6G1},~\cite{Near_ISAC1}. While providing independent directional beams for communication users, integrated waveform design facilitates the detection of sensing targets~\cite{Near_ISAC2}. Due to the extensive use of phase-shifting components, conventional large-scale phased array based ISAC systems may face substantial cost overhead and complex parallel feeding networks~\cite{Near_ISAC3}. To realize low-cost, easily-integrable, and large-scale array assisted ISAC, reconfigurable holographic surfaces~(RHS) present a promising solution. Specifically, an RHS is a serially-fed leaky-wave antenna composed of a large number of low-cost reconfigurable radiating elements. The reference wave injected from the feed sequentially excites each element, where the amplitude of the radiated electromagnetic~(EM) wave is controlled by a diode circuit. By manipulating the states of these diode circuits, the radiated EM waves from each element can be adjusted to form holographic directional beams~\cite{RHS1}.
Based on PCB fabrication, the antenna elements, propagation medium, and feeds of the RHS are lightweight-integrated, enabling direct deployment as transceiving antennas.

When the reference wave excites each RHS element, the EM phase is determined by the distance from the feed, while the amplitude response is controlled by the individual element. Hence, the RHS is typically connected after the RF chain to realize holographic beamforming~\cite{RHS4}. The recently proposed tri-hybrid architecture~\cite{Tri1}, which places reconfigurable antennas sequentially after digital and analog beamformers as an EM layer~\cite{Tri2}, offers new possibilities for holographic ISAC. This structure not only facilitates low-cost large-scale array based ISAC, but also introduces an additional degree of freedom of EM phase control for holographic ISAC. Specifically, each phase shifter in the analog beamformer drives an RHS antenna as a subarray, providing the subarray-level phase shifting and element-level amplitude control capabilities for signal processing. Unlike conventional reflective metasurfaces, RF signals can be directly fed into the metasurface via feeds without additional path loss~\cite{RHS3},~\cite{RIS}, enabling straight integration with existing hybrid beamformers.

Existing works have investigated the tri-hybrid beamforming architecture and its communication applications. To realize energy-efficient large-scale arrays, low-cost and low-power reconfigurable antennas are introduced as an EM layer beyond conventional digital and analog layers~\cite{Tri1},~\cite{Tri2}. In multi-user communications, pinching-antenna-enabled tri-hybrid beamforming is shown to outperform conventional hybrid beamforming with the same number of RF chains~\cite{Pinching}. In~\cite{ERA}, radiation-pattern-reconfigurable antennas are implemented in the EM layer, enhancing the multi-user sum rate through tri-hybrid precoding.
Spectral efficiency maximization is achieved in tri-hybrid beamforming systems via dynamic radiation-center selection among EM-layer elements~\cite{RC}.

In this work, we investigate RHS enabled tri-hybrid holographic ISAC system where the RHSs act as the EM layer.
Although the tri-hybrid architecture facilitates the deployment of low-cost large-scale ISAC, it faces the following challenges compared to conventional hybrid beamforming based systems. \emph{First}, simultaneously multi-user communication and multi-target sensing through joint manipulation of subarray-level phases and element-level radiation amplitudes is non-trivial, due to the escalated computational complexity and distinct feasible domains.
Specifically, the feasible region of the phase control in the analog layer lies within the complex domain, while the radiation amplitude of the RHS elements is constrained to the real domain.
\emph{Second}, due to coupled beamformers in tri-hybrid ISAC, the impact of introducing phase shifters in the analog layer and the effect of the number of antenna elements on holographic ISAC performance remains difficult to characterize theoretically.
To address these challenges, the main contributions of this work are summarized as follows.
\begin{itemize}
\item The tri-hybrid beamformer is employed to realize the low-cost large-scale holographic ISAC paradigm, where a small number of phase shifters in the analog layer provides an additional degree of freedom for subarray-level phase control in holographic beamforming.
For simultaneous multi-user communication and multi-target sensing, a tri-hybrid holographic beamforming problem is formulated to minimize the sensing waveform error while satisfying the minimum user rate requirement. Considering the distinct phase and amplitude constraints in the analog and EM layers, a joint optimization scheme is developed to control subarray-level phases and element-level amplitude responses, updating both beamformers coordinately along the descending direction of the objective function.

\item Enabling by coherent superposition of holographic beams from different subarrays, the beamforming gain enhancement through the incorporation of the analog-layer phase shifters is theoretically demonstrated. Moreover, the impact of the number of RHS elements on holographic ISAC is analytically derived.
The distribution of the optimized RHS element amplitude responses is theoretically characterized, revealing that these responses predominantly cluster near the boundary values. This implies that 1-bit amplitude control can be employed, thereby further reducing both hardware and algorithmic complexity.

\item   Simulation results demonstrate that the holographic beam pattern of the proposed tri-hybrid ISAC scheme enables integrated multi-user communication and multi-target sensing with a controllable trade-off. 
The measured holographic beam gain of the RHS validates the performance improvement of holographic beamforming through subarray-level phase shifting.
Moreover, as the number of RHS elements increases, the proposed scheme approaches and outperforms the hybrid beamforming based ISAC while utilizing significantly fewer phase shifters.

\end{itemize}

\emph{Organization}: The rest of the paper is organized as follows. Section~\ref{Sec2} introduces the RHS model and the RHS-enabled tri-hybrid ISAC system. The optimization problem for RHS-enabled tri-hybrid ISAC and the corresponding iterative algorithm are proposed in Section~\ref{Sec3}. Section~\ref{Sec4} provides the theoretical analysis of the impact of analog-layer phase shifters and the number of EM-layer elements on holographic beamforming gain, along with the distribution of optimized RHS amplitude responses. Simulation results and experimental validation of the proposed tri-hybrid ISAC scheme are presented in Section~\ref{Sec5}, followed by conclusions in Section~\ref{Sec6}.

\emph{Notation}: A scalar is denoted by the unbolded symbol $b$, while a vector is represented by the lowercase bolded symbol $\mathbf{b}$, and an uppercase bolded letter $\mathbf{B}$ stands for a matrix. A set is indicated using the script letter $\mathcal{B}$. The entry located in the $m$-th row and $n$-th column of matrix $\mathbf{B}$ is expressed as $[\mathbf{B}]_{m,n}$, and $\mathbf{B}^H$ denotes the conjugate transpose operation applied to a matrix.
For a complex number $a$, its real component is $\Re\{a\}$ and its imaginary component is $\Im\{a\}$, respectively. The modulus of the complex number $a$ is represented as $|a|$, and the Euclidean norm of a vector \(\mathbf{a}\) is denoted as \(\|\mathbf{a}\|_2\). The Hadamard product is denoted by the operator $\odot$.

\section{System Model}\label{Sec2}

In this section, the RHS-enabled tri-hybrid ISAC system is first introduced, followed by the RHS model. Signal models for communication and sensing are then presented.

\begin{figure}[t]
\setlength{\abovecaptionskip}{0pt}
\setlength{\belowcaptionskip}{0pt}
	\centering
    \includegraphics[width=0.45\textwidth]{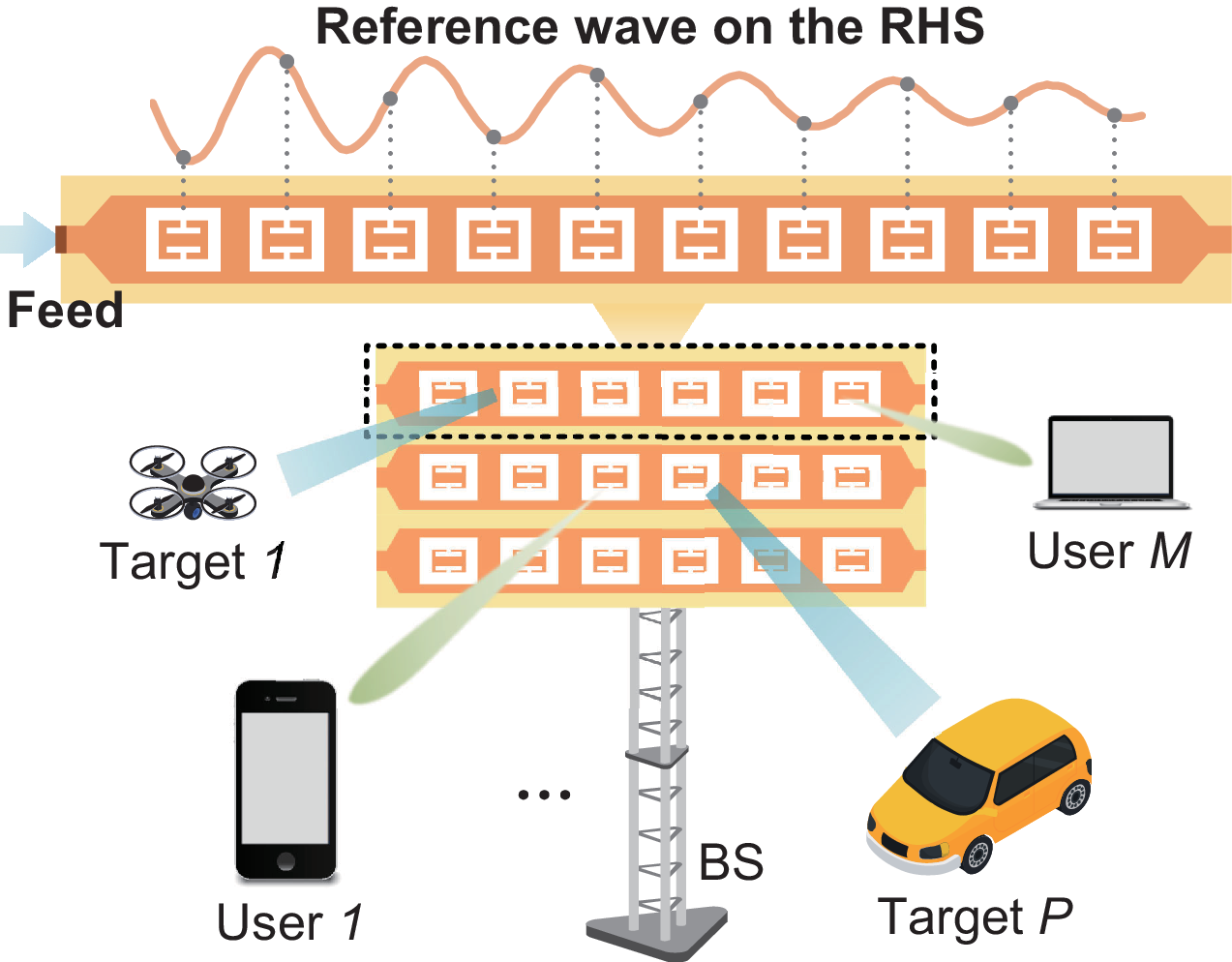}
	\caption{RHS-enabled holographic ISAC system.}
	\label{Fig:BS}
\vspace{-0em}
\end{figure}

\subsection{Scenario Description}
As illustrated in Fig.~\ref{Fig:BS}, to support multi-user communication and multi-target sensing, we consider an ISAC system assisted by a large-scale array. In this system, the BS serves $M$ single-antenna users while simultaneously detecting $P$ sensing targets. Conventional large-scale phased arrays with hybrid beamforming architectures suffer from the requirement for a vast number of phase shifters~\cite{cost}. To address this issue, building upon the hybrid beamforming structure, we employ the low-cost and energy-efficient RHS as the BS transmitting antenna at the EM level to realize tri-hybrid beamforming. This approach avoids the use of numerous phase shifters.
Moreover, the serial feeding mechanism of the RHS enables a single feed to drive a large number of elements, eliminating the individual feeding networks required for each element in phased arrays, thereby enabling the implementation of large-scale ISAC.

\begin{figure}[t]
\setlength{\abovecaptionskip}{0pt}
\setlength{\belowcaptionskip}{0pt}
	\centering
    \includegraphics[width=0.47\textwidth]{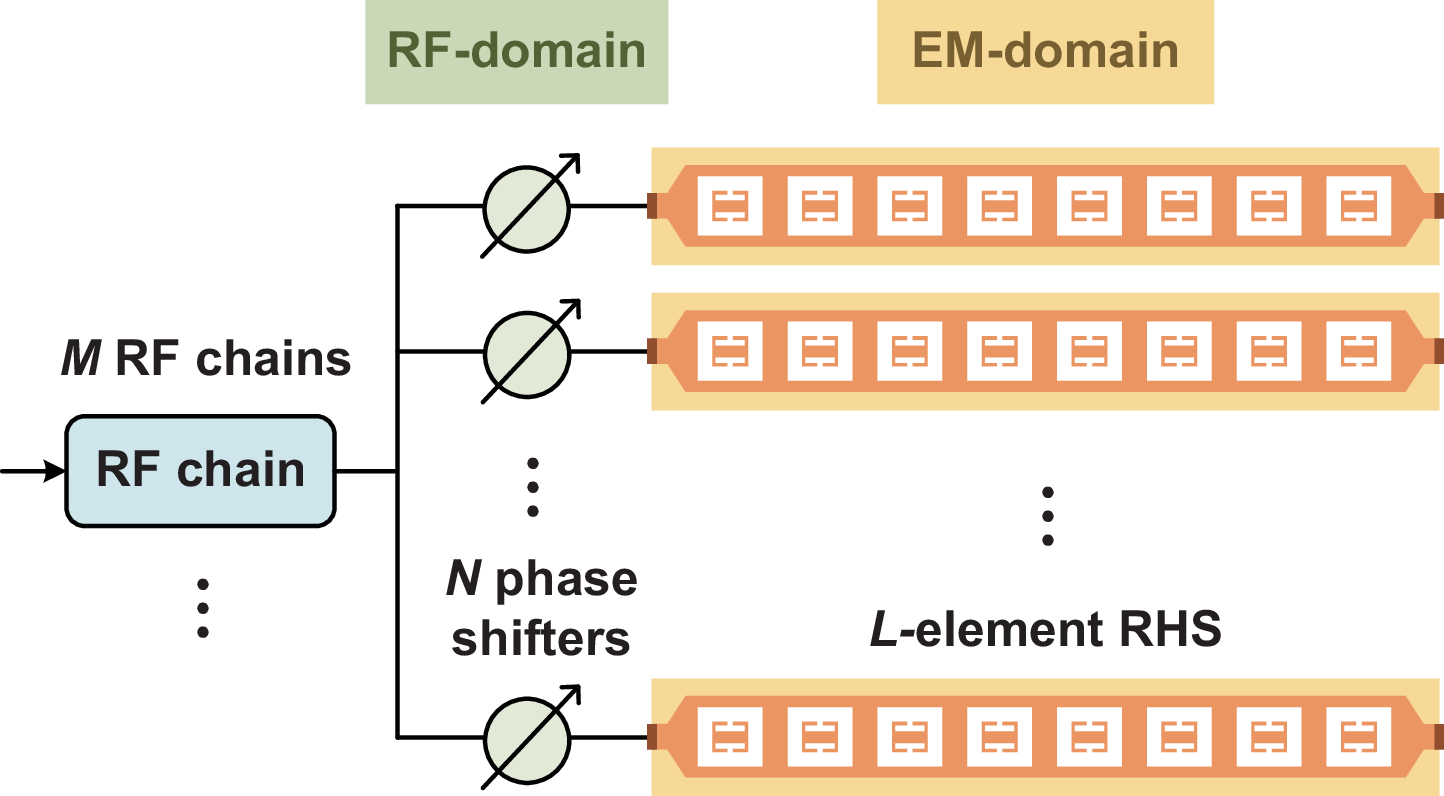}
	\caption{Tri-hybrid structural diagram of a single RF chain.}
	\label{Fig:Subarray}
\vspace{-0em}
\end{figure}

To provide independent data streams for each user, the number of RF chains is assumed to be equal to the number of users, $M$. A partially-connected structure is adopted between the digital and analog beamformers to avoid the extensive use of phase shifters at the analog layer~\cite{Partially}, meaning each RF chain is exclusively connected to a dedicated set of $N$ phase shifters. As shown in Fig.~\ref{Fig:Subarray}, to enable large-scale arrays and further enhance signal processing capabilities, each phase shifter is followed by an RHS comprising $L$ elements, forming one subarray.
Notably, no additional RF chains are allocated for the sensing tasks, as the sensing waveform design is fully integrated into communication resources. Moreover, the tri-hybrid beamforming architecture exhibits inherent coupling among its three layers, which poses significant challenges for the ISAC waveform design.

\subsection{RHS Model}
The RHS operates as a leaky-wave antenna, where the RF signal delivered into the surface via the feed propagates as a reference wave~\cite{RHS1}. As this reference wave successively excites each element, the radiated energy into free space is determined by the radiation amplitude response of the individual element~\cite{Fluid}. By controlling the amplitude response of each element, the coherent superposition of the EM waves radiated from all elements forms a directional holographic beam. The controllable radiation amplitude of each RHS element relies on a complementary electric-LC (cELC) resonator loaded with PIN diodes~\cite{Leaky}. Adjusting the state of the diode circuit enables dynamic regulation of the mutual inductance of the cELC resonator and consequently controls the radiated power from the reference wave. The operational state of the circuit is reconfigured by controlling the bias voltage applied to each diode~\cite{RHS3}.

The reference wave continuously radiates energy outward during propagation, and its phase also changes with propagation. Consider a one-dimensional RHS with $L$ elements, where the feed is placed on the side of the metasurface. The EM wave radiated by the $l$-th element is determined by the reference wave and the radiation amplitude of the element, expressed as~\cite{RHS1},cite{RHS4}
\begin{equation}
E_l = a_{l}\cdot w_{l}\cdot e^{-j\mathbf{k}_s\cdot\mathbf{r}_{l}},
\end{equation}
where $a_{l}$ represents the normalized amplitude response within the range of $[0,1]$, which is controlled by the state of the diode circuit, $w_{l}$ is an energy coefficient that depends on the element's position and radiation efficiency~\cite{TCCN}. The symbol $\mathbf{k}_s$ denotes the wave number on the metasurface, and $\mathbf{r}_{l}$ represents the position vector from the element to the feed.

\subsection{Signal Model}

In the partially-connected architecture, there exist $NM$ phase shifters in the analog beamformer, each driving an $L$-element RHS that constitutes a subarray. The overall transmit antenna is modeled as a uniform planar array~(UPA) defined by $NM$ and $L$ elements in the horizontal and vertical dimensions, respectively. The steering vector $\mathbf{a}(\theta, \phi)$ corresponding to elevation angle $\theta$ and azimuth angle $\phi$ can be expressed as~\cite{channel2}
\begin{gather}
\mathbf{a}(\theta, \phi) = \mathbf{a}_{\text{h}}(\theta, \phi) \otimes \mathbf{a}_{\text{v}}(\theta, \phi), \\[1ex]
\scalebox{0.95}{$\mathbf{a}_h(\theta, \phi)=\frac{1}{\sqrt{NM}}[1,e^{j\frac{2\pi}{\lambda} d_x\sin\theta \cos\phi},\ldots,e^{j\frac{2\pi}{\lambda} d_x(NM-1)\sin\theta \cos\phi}]^T$},\nonumber\\
\scalebox{0.98}{$\mathbf{a}_v(\theta, \phi)=\frac{1}{\sqrt{L}}[1,e^{j\frac{2\pi}{\lambda} d_y\sin\theta \sin\phi},\ldots,e^{j\frac{2\pi}{\lambda} d_y(L-1)\sin\theta \sin\phi}]^T$},\nonumber
\end{gather}
where the symbol $\otimes$ denotes the Kronecker product, and $d_x$ and $d_y$ represent the inter-element spacing in the horizontal and vertical directions, respectively.

The digital and analog beamformers are represented by $\mathbf{F}_D\in\mathbb{C}^{M\times M}$ and $\mathbf{F}_A\in\mathbb{C}^{NM\times M}$, respectively.
The analog beamformer $\mathbf{F}_A$ exhibits a block-diagonal structure, expressed as
\begin{equation}
\mathbf{F}_A = \text{blkdiag}\{\mathbf{f}_{A,1},\mathbf{f}_{A,2},\ldots, \mathbf{f}_{A,M}\},
\end{equation}
where each entry of $\mathbf{f}_{A,m}\in\mathbb{C}^{N\times 1}$ satisfies the constant modulus constraint, i.e., $|\mathbf{f}_{A,m}(n)| = 1, \forall m,n$.
Based on the subarray structure, the EM-layer RHS beamformer $\mathbf{F}_{E}\in\mathbb{C}^{L\times NM}$ also exhibits a block-diagonal form, expressed as
\begin{equation}
\mathbf{F}_{E} = \text{blkdiag}\left\{\mathbf{f}_{E,1} \circ \mathbf{c}_{1}, \ldots, \mathbf{f}_{E,NM} \circ \mathbf{c}_{NM}\right\},
\end{equation}
where $\mathbf{f}_{E,n}=[a_1^n,\ldots,a_L^n]^T$ denotes the amplitude response vector of the $n$-th RHS, and $\mathbf{c}_n=[w_{1}^n\cdot e^{-j\mathbf{k}_s\cdot\mathbf{r}_{1}^n},\ldots,w_{L}^n\cdot e^{-j\mathbf{k}_s\cdot\mathbf{r}_{L}^n}]$ is the corresponding coefficient vector.

\textbf{Communication Model:}
Based on the Saleh-Valenzuela channel model, the channel of the $m$-th user can be expressed as~\cite{channel2}
\begin{equation}
\mathbf{h}_m=\sqrt{\frac{NML}{Z_m}}\sum_{z=1}^{Z_m} \beta_z^{m}\mathbf{a}(\theta_z^{m}, \phi_z^{m}),
\end{equation} 
where $Z_m$ is the number of paths, $\beta_z^{m}$ denotes the complex gain of $z$-th path, and $(\theta_z^{m}, \phi_z^{m})$ correspond to the elevation and azimuth angles of departure for the path, respectively.
The received signal of the $m$-th user can be represented as
\begin{equation}
y_m=\mathbf{h}_{m}^H\mathbf{F}_{E}\mathbf{F}_A\mathbf{f}_{D,m}s_m+\sum_{{m'}\neq m}\mathbf{h}_{m}^H\mathbf{F}_{E}\mathbf{F}_A\mathbf{f}_{D,m'}s_{m'}+n_m,
\nonumber
\end{equation}
where $\mathbf{f}_{D,m}$ represents the $m$-th column of digital beamformer $\mathbf{F}_{D}$, $s_m$ is the transmitted signal for the $m$-th user satisfying $\mathbb{E}[|s_m|^2]=1$, and $n_m\sim\mathcal{CN}(0,\sigma^2)$ represents the additive white Gaussian noise.
Hence the data rate of user $m$ can be given by\cite{ISAC_C}
\begin{equation}
R_m= \log_2\left(1+\frac{|\mathbf{h}_{m}^H\mathbf{F}_{E}\mathbf{F}_A\mathbf{f}_{D,m}|^2}{\sum_{{m'}\neq m}|\mathbf{h}_{m}^H\mathbf{F}_{E}\mathbf{F}_A\mathbf{f}_{D,m'}|^2+\sigma^2}\right).
\end{equation}

\textbf{Sensing Model:}
No dedicated RF chain is allocated for the sensing function, and its waveform design is integrated within the communication resources. By optimizing the tri-hybrid beamformers $\mathbf{F}_D$, $\mathbf{F}_A$, and $\mathbf{F}_E$, the beamforming gain toward the sensing direction $(\theta, \phi)$ is expressed as~\cite{ISAC_S}
\begin{equation}
G(\theta, \phi)=|\mathbf{a}^H(\theta, \phi)\mathbf{w}_{\text{sense}}|^2,
\end{equation}
where the beamformer $\mathbf{w}_{\text{sense}}=\sum_{m=1}^M\mathbf{F}_{E}\mathbf{F}_A\mathbf{f}_{D,m}$.

\section{RHS-Enabled Tri-Hybrid Holographic ISAC}\label{Sec3}

In this section, the optimization objective for RHS-enabled tri-hybrid holographic ISAC is formulated and subsequently transformed. The optimization methods for beamformers in the digital, analog, and EM layers are then presented.

\subsection{Problem Formulation and Transformation}

Given $P$ sensing directions and without dedicating additional RF chains to sensing, the ISAC problem is formulated as the design of a sensing waveform through the optimization of the tri-hybrid beamformer, subject to the constraint of guaranteeing the user data rate~\cite{ISAC_S},~\cite{ISAC_Tut}, expressed as
\begin{align}
\min_{\mathbf{F}_D, \mathbf{F}_A, \mathbf{F}_E} \quad & \left\|\mathbf{g}_{\text{sense}}- \mathbf{b}\right\|_2  \label{p1}\\
\text{s.t.} \quad & R_m \geq R_{\text{th}}, \quad \forall m = 1,\ldots,M  \label{c1} \\
& \|\mathbf{F}_A \mathbf{F}_D\|_F^2 \leq P_{\text{max}} \label{c2} \\
& |\mathbf{f}_{A,m}(n)| = 1, \forall m,n \label{c3} \\
& 0 \leq \mathbf{f}_{E,n}(l) \leq 1, \quad \forall n,l,  \label{c4}
\end{align}
where $\mathbf{g}_{\text{sense}} = [G(\theta_1, \phi_1), \cdots,G(\theta_P, \phi_P)]$ represents the beamforming gain for each direction $(\theta_p, \phi_p)$, and $\mathbf{b}=[b_1,\cdots,b_P]$ represents the desired beamforming gain for each direction.
Constraint~(\ref{c1}) ensures that the achievable rate for each user meets the threshold $R_{\text{th}}$. Constraint~(\ref{c2}) imposes the power constraints on the digital and analog beamformers, noting that the RHS operates passively by receiving RF signals without additional power consumption. Constraint~(\ref{c3}) represents the constant modulus constraint for the elements of the analog beamformer. Constraint~(\ref{c4}) defines the range of the normalized amplitude response for the RHS elements.

Owing to the non-convexity of the rate constraint~(\ref{c1}), Problem~(\ref{p1}) is non-convex and thus directly solving it is intractable.
By incorporating this constraint into the objective function, the problem is transformed into
\begin{align}
\min_{\mathbf{F}_D, \mathbf{F}_A, \mathbf{F}_E} \quad &\left\|\mathbf{g}_{\text{sense}}- \mathbf{b}\right\|_2 + \mu \sum_{m=1}^M \max(0, R_{\text{th}} - R_m)^2, \label{P2}\\
\text{s.t.} \quad & \text{   }\text{   } \text{(\ref{c2}) - (\ref{c4})} ,\quad  \nonumber
\end{align}
where $\mu$ represents an adaptive penalty coefficient, which increases when the user rate requirement is not met and decreases when the requirement is satisfied.
Due to the coupling among the tri-hybrid beamformers, an exhaustive simultaneous search would incur high computational complexity. Therefore, we adopt an iterative approach to optimize the beamformers sequentially. Specifically, in the $(t+1)$-th iteration, the penalty coefficient is updated according to the following rule:
\begin{equation}
\mu^{(t+1)} = 
\begin{cases}
\min(\rho \mu^{(t)}, \mu_{\max}), & \text{if } \min_m R_m < R_{\text{th}}, \\
\max(\mu^{(t)}/\rho, \mu_{\min}), & \text{else},
\end{cases}
\end{equation}
where $\rho$ denotes a multiplicative factor, and the penalty coefficient is constrained within the interval $[\mu_{\min},\mu_{\max}]$.

\subsection{Digital Beamformer Optimization}\label{III-A}
During each iteration, the tri-hybrid beamformers are alternately optimized to solve Problem~(\ref{P2}) due to the coupling among them. For the digital beamformer $\mathbf{F}_D$, given the analog beamformer and the RHS beamformer, it is optimized using the sequential quadratic programming~(SQP) algorithm, considering its power constraints and the differentiability of the objective function~\cite{SQP}.
To facilitate numerical optimization, the real and imaginary parts of each element in $\mathbf{F}_D$ are combined into a real-valued vector $\mathbf{x}$, expressed as
\begin{equation}
\mathbf{x} = \begin{bmatrix}
\text{vec}(\Re\{\mathbf{F}_D\}) \\
\text{vec}(\Im\{\mathbf{F}_D\})
\end{bmatrix}.
\end{equation}
The corresponding optimization problem in the $t$-th iteration is reformulated as
\begin{equation}
\begin{aligned}
\min_{\mathbf{x}} \quad & \mathcal{F}(\mathbf{x}) = \mathcal{J}_{\text{sense}}(\mathbf{x}) + \mu^{(t)} \cdot \mathcal{V}_{\text{rate}}(\mathbf{x}) \\
\text{s.t.} \quad & \mathcal{P}(\mathbf{x}) = \|\mathbf{F}_A \mathbf{F}_D(\mathbf{x})\|_F^2 - P_{\text{max}} \leq 0,
\end{aligned}
\end{equation}
where $\mathcal{J}_{\text{sense}}=\left\|\mathbf{g}_{\text{sense}}- \mathbf{b}\right\|_2$ and $\mathcal{V}_{\text{rate}}(\mathbf{x})=\sum_{m=1}^M \max(0, R_{\text{th}} - R_m)^2$.
To update $\mathbf{x}$, the gradient of the objective function needs to be obtained.
Based on the Wirtinger derivative, the gradient of the sensing error term with respect to $\mathbf{F}_D^*$ is given by
\begin{align}
\nabla^* \mathcal{J}_{\text{sense}} &= \sum_{p=1}^{P} 2(G_p - b_p) \cdot \frac{\partial G_p}{\partial \mathbf{F}_D^*}, \\
\frac{\partial G_p}{\partial \mathbf{F}_D^*} &=  \mathbf{F}_A^H \mathbf{F}_E^H \mathbf{a}_p \mathbf{a}_p^H \mathbf{F}_E\mathbf{F}_A \sum_{m=1}^M\mathbf{f}_{D,m} \mathbf{1}_M^T ,
\end{align}
where $\mathbf{1}_M \in \mathbb{R}^{m}$ is an all-ones vector.
Similarly, the gradients of the rate term are given by
\begin{align}
\nabla^* \mathcal{V}_{\text{rate}} &= -2 \sum_{m=1}^{M} v_m \cdot \frac{\partial R_m}{\partial \mathbf{F}_D^*},\\
\frac{\partial R_m}{\partial \mathbf{F}_D^*} &= \frac{1}{\ln 2} \cdot \frac{1}{1 + \text{SINR}_m} \cdot \frac{\partial \text{SINR}_m}{\partial \mathbf{F}_D^*},
\end{align}
where $v_m = \max(0, R_{\text{th}} - R_m)$, and $\text{SINR}_m= \frac{|\mathbf{h}_{m}^H\mathbf{F}_{E}\mathbf{F}_A\mathbf{f}_{D,m}|^2}{\sum_{{m'}\neq m}|\mathbf{h}_{m}^H\mathbf{F}_{E}\mathbf{F}_A\mathbf{f}_{D,m'}|^2+\sigma^2}$.
Therefore, the total gradient is expressed as
\begin{equation}
\nabla \mathcal{F}(\mathbf{x}) = 
\begin{bmatrix}
\Re\{\text{vec}(\nabla \mathcal{J}_{\text{sense}})\} \\
\Im\{\text{vec}(\nabla \mathcal{J}_{\text{sense}})\}
\end{bmatrix}
+ \mu^{(t)} \cdot 
\begin{bmatrix}
\Re\{\text{vec}(\nabla \mathcal{V}_{\text{rate}})\} \\
\Im\{\text{vec}(\nabla \mathcal{V}_{\text{rate}})\}
\end{bmatrix}.\nonumber
\end{equation}
Detailed derivations of the gradients are provided in {Appendix~\ref{app_1}}.
To determine the update direction $\mathbf{d}$ for $\mathbf{x}$, in the $n$-th inner loop, the following quadratic programming subproblem is solved
\begin{equation}
\begin{aligned}
\min_{\mathbf{d}} \quad & \frac{1}{2}\mathbf{d}^T \mathbf{B}^{(n)} \mathbf{d} + (\nabla\mathcal{F}(\mathbf{x}^{(n)}))^T \mathbf{d} \\
\text{s.t.} \quad & (\nabla\mathcal{P}(\mathbf{x}^{(n)}))^T \mathbf{d} + \mathcal{P}(\mathbf{x}^{(n)}) \leq 0,
\end{aligned}
\end{equation}
where $\mathbf{B}$ is the Hessian approximation matrix. $\mathcal{P}(\mathbf{x})$ is the gradient of the constraint function, and its real and imaginary components are $2\Re\left\{\text{vec}\left( \mathbf{F}_A^H \mathbf{F}_A \mathbf{F}_D \right)\right\}$ and $2\Im\left\{\text{vec}\left( \mathbf{F}_A^H \mathbf{F}_A \mathbf{F}_D \right)\right\}$, respectively.
The KKT conditions and the solutions for this quadratic program are given by
\begin{equation}
\begin{bmatrix}
\mathbf{B}^{(n)} & \nabla\mathcal{P}(\mathbf{x}^{(n)}) \\
(\nabla\mathcal{P}(\mathbf{x}^{(n)}))^T & 0
\end{bmatrix}
\begin{bmatrix}
\mathbf{d}^* \\ \lambda^*
\end{bmatrix}
= 
\begin{bmatrix}
-\nabla\mathcal{F}(\mathbf{x}^{(n)}) \\ -\mathcal{P}(\mathbf{x}^{(n)})
\end{bmatrix},
\end{equation}
\vspace{-0.8em}
\begin{gather}
\lambda^* = \frac{(\nabla\mathcal{P}(\mathbf{x}^{(n)}))^T (\mathbf{B}^{(n)})^{-1} \nabla\mathcal{F}(\mathbf{x}^{(n)}) -\mathcal{P}(\mathbf{x}^{(n)})}{(\nabla\mathcal{P}(\mathbf{x}^{(n)}))^T (\mathbf{B}^{(n)})^{-1} \nabla\mathcal{P}(\mathbf{x}^{(n)})}, \\[1ex]
\mathbf{d}^* = -(\mathbf{B}^{(n)})^{-1} (\nabla\mathcal{F}(\mathbf{x}^{(n)}) + \lambda^* \nabla\mathcal{P}(\mathbf{x}^{(n)})).
\end{gather}
Following this, the vector $\mathbf{x}$ is updated as
\begin{gather}
\mathbf{x}^{(n+1)} = \mathbf{x}^{(n)} + \alpha^{(n)} \mathbf{d}^*,
\end{gather}
where the step size $\alpha^{(n)}$ is determined based on the Armijo condition.
The Hessian approximation matrix $\mathbf{B}$ is subsequently updated using the BFGS method~\cite{BFGS}, the details of which are omitted here due to its general applicability.

\subsection{Joint Optimization of the Analog and RHS Beamformers}
We now present the joint optimization approach for the analog and RHS beamformers, where the elements in $\mathbf{F}_A$ implement subarray-level phase control while the elements in $\mathbf{F}_E$ optimize the corresponding amplitude response.
Considering the constant modulus constraints of the elements in $\mathbf{F}_A$ and the controllable amplitude response range of the elements in $\mathbf{F}_E$, an adaptive gradient projection method is employed to optimize these two beamformers. Specifically, the gradient of the objective function is first computed, and the updated variables are then projected onto the feasible region according to the gradient.
The corresponding optimization problem in the $t$-th iteration is reformulated as
\begin{align}
\min_{\mathbf{F}_A,\mathbf{F}_E} \quad & \mathcal{F}(\mathbf{F}_A,\mathbf{F}_E) = \mathcal{J}_{\text{sense}}(\mathbf{F}_A,\mathbf{F}_E) + \mu^{(t)}\cdot \mathcal{V}_{\text{rate}}(\mathbf{F}_A,\mathbf{F}_E)  \nonumber\\
\text{s.t.} \quad & |\mathbf{f}_{A,m}(n)| = 1, \forall m,n   \\
& 0 \leq \mathbf{f}_{E,n}(l) \leq 1, \quad \forall n,l.  \nonumber
\end{align}
During the inner iterations of the joint optimization, we alternately update $\mathbf{F}_E$ and $\mathbf{F}_A$.
To minimize the objective function, the amplitude responses of the RHS are optimized, while the elements in $\mathbf{F}_A$ are updated to control the phase of corresponding RHS subarrays.
Using the chain rule, the gradient of the sensing term with respect to $\mathbf{F}_E$ is given by
\begin{align}
\frac{\partial \mathcal{J}_{\text{sense}}}{\partial \mathbf{F}_{E,\text{amp}}} &= \sum_{p=1}^P 2\delta_p \cdot \frac{\partial G_p}{\partial \mathbf{F}_{E,\text{amp}}},\\
\frac{\partial G_p}{\partial \mathbf{F}_{E,\text{amp}}} &= 2 \cdot \Re\left\{ \mathbf{C}_{\text{phase}}^* \odot \left[ \mathbf{a}_p \mathbf{a}_p^H \mathbf{G} \mathbf{f}_{{D}} \mathbf{f}_{{D}}^H \mathbf{F}_A^H \right] \right\},
\end{align}
where $\mathbf{F}_{E,\text{amp}}=\text{blkdiag}\left\{\mathbf{f}_{E,1}, \ldots, \mathbf{f}_{E,NM}\right\}$ and $\mathbf{C}_{\text{phase}}=\text{blkdiag}\left\{\mathbf{c}_{1}, \ldots, \mathbf{c}_{NM}\right\}$ represent the amplitude and phase matrices of the RHS elements, respectively.
Meanwhile, the gradient of the rate term with respect to $\mathbf{F}_E$ is calculated as
\begin{align}
\frac{\partial \mathcal{V}_{\text{rate}}}{\partial \mathbf{F}_{E,\text{amp}}} &= \sum_{m=1}^M -2v_m \cdot \frac{\partial R_m}{\partial \mathbf{F}_{E,\text{amp}}},\\
\frac{\partial R_m}{\partial \mathbf{F}_{E,\text{amp}}} &= \frac{1}{\ln 2} \cdot \frac{1}{1 + \text{SINR}_m} \cdot \frac{\partial \text{SINR}_m}{\partial \mathbf{F}_{E,\text{amp}}}.
\end{align}
Combining the gradients from the sensing and rate terms yields the total gradient for $\mathbf{F}_{E,\text{amp}}$ as $\nabla \mathcal{F}(\mathbf{F}_{E,\text{amp}})$.
During the $n$-th inner iteration of the joint optimization, an adaptive step size $\epsilon^{(n)}$ is subsequently employed to project the updated $\mathbf{F}_{E,\text{amp}}$ onto the feasible region:
\begin{gather}
\mathbf{F}_{E,\text{amp}}^{(n+1)} = \mathcal{P}_{\mathcal{E}} \left( \mathbf{F}_{E,\text{amp}}^{(n)} - \epsilon^{(n)} \nabla \mathcal{F}(\mathbf{F}^{(n)}_{E,\text{amp}}) \right), \\
\mathcal{P}_{\mathcal{E}}(\mathbf{X}) = \max(\mathbf{0}, \min(\mathbf{1}, \mathbf{X})),
\end{gather}
where $\mathcal{P}_{\mathcal{E}}$ denotes the projection operator for $\mathbf{F}_{E,\text{amp}}$.

\begin{algorithm}[t]
\caption{Tri-Hybrid Beamformer Optimization}
\label{alg:tri_hybrid}
\begin{algorithmic}[1]
\STATE \textbf{Input:} $\mathbf{b}, R_{th}, \mu^{(0)}, P_{max}, \mathbf{h}_m, \mathbf{a}_p, F_D^0, F_A^0, F_E^0$
\FOR{$t = 1$ \textbf{to} $T_{max}$}
    \STATE Update penalty coefficient:
    \STATE \hspace{0.5cm} $\mu^{(t)} = 
    \begin{cases}
    \min(\rho \mu^{(t-1)}, \mu_{\max}),& \text{if } \min_m R_m < R_{\text{th}}, \\
    \max(\mu^{(t-1)}/\rho, \mu_{\min}),& \text{else},
    \end{cases}
    $
    
    \FOR{$n = 1$ \textbf{to} $I_{inner}$}
        \STATE $\mathbf{d}^* = -(\mathbf{B}^{(n)})^{-1} (\nabla\mathcal{F}(\mathbf{x}^{(n)}) + \lambda^* \nabla\mathcal{P}(\mathbf{x}^{(n)}))$
        \STATE $\mathbf{x}^{(n+1)} = \mathbf{x}^{(n)} + \alpha^{(n)} \mathbf{d}^*$
    \ENDFOR
    \STATE $F_D^{(t)} \gets \mathbf{x}^{(I_{inner})}$
    
    \FOR{$m = 1$ \textbf{to} $J_{inner}$}
        \STATE Compute $\nabla \mathcal{F}(\mathbf{F}^{(n)}_{E,\text{amp}})$ and $\nabla \mathcal{F}(\mathbf{F}^{(n)}_{A})$
        \STATE $\mathbf{F}_{E,\text{amp}}^{(n+1)} = \mathcal{P}_{\mathcal{E}} \left( \mathbf{F}_{E,\text{amp}}^{(n)} - \epsilon^{(n)} \nabla \mathcal{F}(\mathbf{F}^{(n)}_{E,\text{amp}}) \right)$
        \STATE ${\mathbf{F}}_A^{(n+1)} = \mathcal{P}_{\mathcal{A}} \left(\mathbf{F}_A^{(n)} - \gamma^{(n)} \nabla \mathcal{F}(\mathbf{F}^{(n)}_A) \right)$
    \ENDFOR
    \STATE $F_E^{(t)} \gets F_E^{J_{inner}}, \quad F_A^{(t)} \gets F_A^{J_{inner}}$
    
    \IF{$\min_m R_m > R_{\text{th}}$ \AND $\Delta_{\mathcal{J}_{\text{sense}}} < \tau_{th}$}
        \STATE \textbf{break}
    \ENDIF
\ENDFOR
\STATE \textbf{Output:} $F_D, F_A, F_E$
\end{algorithmic}
\end{algorithm}

Subsequently, to further reduce the value of the objective function, each element in $\mathbf{F}_{A}$ is adjusted to collectively shift the phase of its corresponding RHS subarray.
Using the chain rule and matrix differentiation, the gradient of the sensing term is expressed as
\begin{align}
\nabla^* \mathcal{J}_{\text{sense}} &= \sum_{p=1}^P 2\delta_p \cdot \frac{\partial G_p}{\partial \mathbf{F}^*_A}, \\
\frac{\partial G_p}{\partial \mathbf{F}^*_A} & =  \mathbf{F}_{E}^H \mathbf{a}_p \mathbf{a}_p^H \mathbf{G} \mathbf{f}_{D} \mathbf{f}_{D}^H.
\end{align}
The gradient of the rate term is expressed as
\begin{align}
\nabla^* \mathcal{V}_{\text{rate}} &= \sum_{m=1}^M -2v_m \cdot \frac{\partial R_m}{\partial \mathbf{F}^*_A},\\
\frac{\partial R_m}{\partial \mathbf{F}^*_A} &= \frac{1}{\ln 2} \cdot \frac{1}{1 + \text{SINR}_m} \cdot \frac{\partial \text{SINR}_m}{\partial \mathbf{F}^*_A}.
\end{align}
Detailed derivations of the gradients for $\mathbf{F}_A$ are provided in {Appendix~\ref{app_2}}.
After obtaining the total gradient $\nabla \mathcal{F}(\mathbf{F}_A)$, the updated $\mathbf{F}_A$ based on the gradient is projected onto the feasible region as
\begin{gather}
{\mathbf{F}}_A^{(n+1)} = \mathcal{P}_{\mathcal{A}} \left(\mathbf{F}_A^{(n)} - \gamma^{(n)} \nabla \mathcal{F}(\mathbf{F}^{(n)}_A) \right), \\
[\mathcal{P}_{\mathcal{A}}(\mathbf{X})]_{m,n} = \frac{[\mathbf{X}]_{m,n}}{|[\mathbf{X}]_{m,n}|}.
\end{gather}
where $\mathcal{P}_{\mathcal{A}}$ denotes the constant modulus projection operator for $\mathbf{F}_A$.
The complete optimization procedure for the tri-hybrid beamformer is summarized in \textbf{Algorithm}~\ref{alg:tri_hybrid}, where $T_{max}$ denotes the maximum number of outer iterations, $I_{inner}$ and $J_{inner}$ represent the maximum numbers of inner iterations, and $\tau_{th}$ is the threshold for sensing beam error.

\section{Performance Analysis of Tri-Hybrid Holographic ISAC}\label{Sec4}

In this section, since ISAC performance depends on beamforming gain, its enhancement through the incorporation of analog-layer phase shifters is theoretically demonstrated.
Subsequently, the impact of the number of RHS elements and the distribution of optimized amplitude responses are analytically characterized. Finally, the computational complexity and convergence properties of the proposed algorithm are discussed.

\subsection{Effect of the Analog Layer on Beamforming Gain}

The RHS enables dynamic control of radiation amplitude while maintaining fixed EM phase at each element~\cite{RHS2}. The introduction of phase shifters in the analog layer of the tri-hybrid beamforming architecture provides subarray-level phase manipulation capability for the RHS. To illustrate this, we theoretically analyze the performance enhancement of holographic beamforming gain through subarray-level phase control in the analog layer.

In the absence of the analog layer, which is equivalent to having fixed element phases in $\mathbf{F}_A$, the beamforming gain for the $p$-th direction is expressed as
\begin{equation}
\bar{G}_p = \left| \sum_{i=1}^{NM} \mathbf{a}_{p,i}^H \mathbf{F}_{E,i} e^{j{\bar\theta_i}} u_i \right|^2,
\end{equation}
where $\mathbf{F}_{E,i} = \text{diag}(a^i_1c^i_1, \cdots, a^i_Lc^i_L)$ denotes the beamformer of the RHS subarray connected to the $i$-th phase shifter, and $\bar\theta_i$ represents the fixed phase of the $i$-th phase shifter.
When the analog layer provides subarray-level phase shifts, the optimal phase $ \theta_i^* = \angle(\mathbf{a}_{p,i}^H \mathbf{F}_{E,i}  u_i) $ enables coherent superposition of signals from all subarrays.
The resulting beamforming gain is given by
\begin{equation}
G_p^* = \left( \sum_{i=1}^{NM} | \mathbf{a}_{p,i}^H \mathbf{F}_{E,i} u_i| \right)^2.
\end{equation}
According to the Cauchy-Schwarz inequality, the introduction of phase shifters in the analog layer enhances the beamforming gain, expressed as
\begin{align}
\Delta G_p^{\text{phase}} &= \left( \sum_{i=1}^{NM} | \mathbf{a}_{p,i}^H \mathbf{F}_{E,i} u_i| \right)^2 - \left| \sum_{i=1}^{NM} \mathbf{a}_{p,i}^H \mathbf{F}_{E,i} e^{j{\bar\theta_i}} u_i \right|^2  \nonumber\\
&\geq \zeta_p \cdot \left( 2\left| \sum_{i=1}^{NM} \mathbf{a}_{p,i}^H \mathbf{F}_{E,i} e^{j{\bar\theta_i}} u_i \right| \right)  \nonumber\\
&= 2\zeta_p \sqrt{\bar{G}_p},
\end{align}
where $\zeta_p = \sum_{i=1}^{NM} |\mathbf{a}_{p,i}^H \mathbf{F}_{E,i} u_i| - \left| \sum_{i=1}^{NM} \mathbf{a}_{p,i}^H \mathbf{F}_{E,i} e^{j{\bar\theta_i}} u_i \right| \geq 0$.
It is noteworthy that even for large-scale tri-hybrid arrays, only a small number of phase shifters is required to enhance the performance of holographic ISAC, since each subarray is equipped with only one phase shifter. For instance, with $M$ = 4, $N$ = 4, and $L$ = 48, merely 16 phase shifters can enable a large-scale holographic beamformer with $16\times 48$ elements.

\begin{remark}\label{remark2}
In tri-hybrid beamforming, the introduction of phase shifters in the analog layer enhances holographic beamforming gain by enabling subarray-level phase manipulation for the RHS.
\end{remark}

\subsection{Impact of the Number of RHS Elements}

As a type of leaky-wave antenna, the reference wave in the RHS gradually attenuates as each element radiates EM energy. Upon reaching a specific element, the reference wave's energy may be entirely depleted. Consequently, investigating the holographic beamforming gain with increasing numbers of RHS elements is crucial.
Next, we theoretically derive the variation in sensing waveform error when the number of RHS elements increases from $L$ to $L+1$.

The sensing beamformer and steering vector for an RHS with $L$ elements are represented as
\begin{equation}
\mathbf{w}_{\text{sense}}(L) = 
\begin{bmatrix}
\mathbf{w}_1(L) \\
\mathbf{w}_2(L) \\
\vdots \\
\mathbf{w}_{NM}(L)
\end{bmatrix},
\mathbf{a}_p(L) = 
\begin{bmatrix}
\mathbf{a}_{p,1}(L) \\
\mathbf{a}_{p,2}(L) \\
\vdots \\
\mathbf{a}_{p,NM}(L)
\end{bmatrix},
\end{equation}
where each sub-block $\mathbf{w}_i(L)$ corresponds to a sub-array of $L$ elements driven by the $i$-th phase shifter, and $\mathbf{a}_{p,i}(L) \in \mathbb{C}^{L}$ denotes the response of the $i$-th sub-array towards the $p$-th direction.
Sub-block $\mathbf{w}_i(L) = u_i \cdot [a^i_1c^i_1, \cdots, a^i_Lc^i_L]^T$, and $u_i$ represents the output signal of the $i$-th phase shifter.
When the number of elements increases to $L+1$, the beamforming gain towards the $p$-th direction is given by
\begin{gather}
\scalebox{0.95}{$G_p(L+1) = \left| \sum_{i=1}^{NM} \left[ \mathbf{a}_{p,i}^H(L), h_{p,i}^*(L+1) \right] \begin{bmatrix} \mathbf{w}_i(L) \\ u_i a^i_{L+1}c^i_{L+1} \end{bmatrix} \right|^2.$} \nonumber
\end{gather}
The incremental beamforming gain relative to the $L$-element case is expressed as
\begin{align}
\Delta G_p(L)&= G_p(L+1) - G_p(L)  \\
&= \left| \xi_p(L) + w_{L+1} \Gamma_p(L+1) \right|^2 - \left| \xi_p(L) \right|^2 \nonumber\\
&= 2w_{L+1} \Re\left\{ \xi_p^*(L) \Gamma_p(L+1) \right\} + w_{L+1}^2 |\Gamma_p(L+1)|^2,\nonumber
\end{align}
where $w_{L+1}$ is the energy coefficient of the $(L+1)$-th RHS element, $\xi_p(L) = \sqrt{G_p(L)}$, $\Gamma_p(L+1) = \sum_{i=1}^{NM} h_{p,i}^* u_i a^i_{L+1} e^{-j\mathbf{k}_s\cdot\mathbf{r}_{L+1}^i}$, and $h_{p,i}$ is the $(L+1)$-th element of $\mathbf{a}_{p,i}(L+1)$.
Considering the serial transmission mode of the RHS, the energy coefficient $w_{L+1} = \sqrt{\eta(1-p_{on}\eta)^L}$ exhibits an exponential decay characteristic, where $p_{on} \in [0,1]$ denotes the radiation probability of the RHS element and $\eta \in [0,1]$ represents the radiation efficiency.
Since the energy coefficient is less than 1 and exhibits exponential decay, there exists a specific number of elements $L^*$, beyond which the gain increment converges to zero as the number of RHS elements increases, expressed as
\begin{equation}
\lim_{L \to L^*} \Delta G_p(L) = 0.
\end{equation}

\begin{remark}\label{remark1}
The holographic beamforming gain of the tri-hybrid ISAC system initially improves and then saturates as the number of RHS elements increases.
\end{remark}

\subsection{Distribution of the Optimized Amplitude Responses}

\begin{figure}[t]
\setlength{\abovecaptionskip}{0pt}
\setlength{\belowcaptionskip}{0pt}
	\centering
    \includegraphics[width=0.5\textwidth]{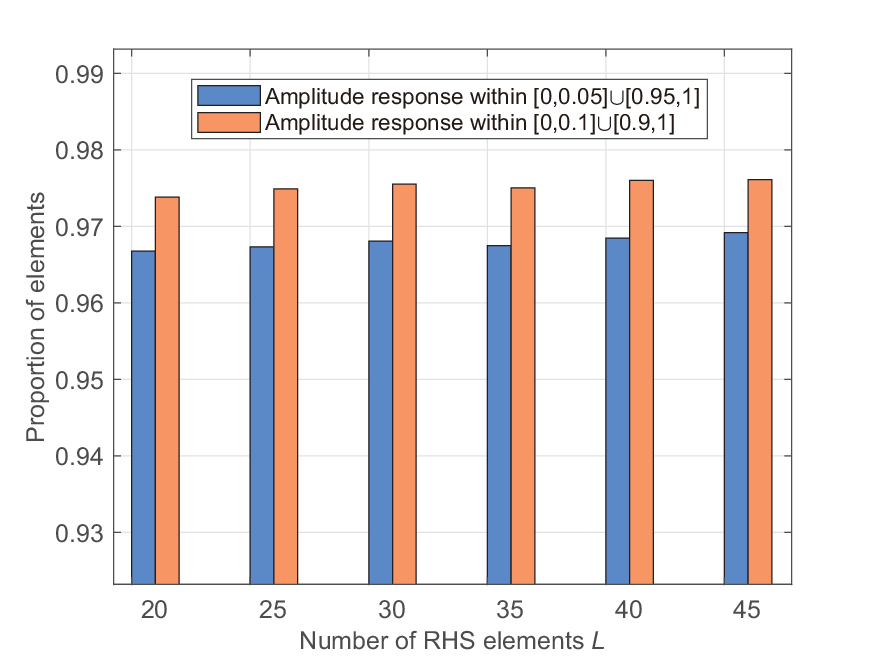}
	\caption{Distribution of the optimized amplitude responses.}
	\label{Fig:proportion}
\vspace{-0em}
\end{figure}

Considering the practical structure of RHS elements, the control of radiation amplitude is not arbitrary but constrained by the number of states available in the diode circuit. In fact, following the tri-hybrid holographic ISAC scheme proposed in Section III, the optimized amplitude responses of most RHS elements are clustered near the boundaries of the feasible region. As shown in Fig.~\ref{Fig:proportion}, as the number of RHS elements increases, the proportion of elements with optimized amplitude responses lying within 5\% and 10\% of the boundary values exceeds 95\% in both cases.

To reveal why most optimized amplitude responses cluster near the boundaries of the feasible region $[0,1]$, we approximate the objective function as a quadratic function of the element-wise amplitude responses. Moreover, we demonstrate that the boundary values~(0/1) can serve as suboptimal solutions for the amplitude responses, thereby reducing the complexity of both algorithmic implementation and hardware realization for holographic ISAC systems.

For the $l$-th RHS element corresponding to the $i$-th phase shifter, the objective function can be approximated as a quadratic function of its amplitude value $a_l^i$, expressed as
\begin{equation}
\mathcal{F}(a_l^i) \approx \mathcal{F}(0) + \mathcal{F}'(0) \cdot a_l^i + \frac{1}{2} \mathcal{F}''(0) \cdot (a_l^i)^2.
\end{equation}
Assuming the interference among communication users is eliminated through the design of the tri-hybrid beamformers, i.e., $\mathbf{h}_m^H \mathbf{F}_E \mathbf{F}_A \mathbf{f}_{D,m'} = 0, \quad \forall m \neq m'$.
Then the constant term can be represented as
\begin{equation}
\mathcal{F}(0) = \sum_{p=1}^{P} (|z_p|^2 - b_p)^2 + \mu \sum_{m=1}^{M} \max(0, R_{\text{th}} - \log_2(1 + \frac{|p_m|^2}{\sigma^2}))^2, \nonumber
\end{equation}
where the notation $z_p = \sum_{(j,k) \neq (i,l)} [\mathbf{a}_{p,j}]_k a^j_{k} c^j_{k} u_j$, 
$p_m = \mathbf{h}_m^H \mathbf{F}_E^{(i,l)}(0) \mathbf{F}_A \mathbf{f}_{D,m}$, and $\mathbf{F}_E^{(i,l)}(0) = \mathbf{F}_E \big|_{a_{l}^i = 0}.$
The coefficient $\mathcal{F}'(0)$ of the linear term can be expressed as
\begin{align}
\mathcal{F}'(0) &= \sum_{p=1}^{P} 4(|z_p|^2 - b_p) \Re\{z_p^* q_p\} \\
&- \mu \sum_{m=1}^{M} \frac{4 v_m}{\ln 2 \cdot \sigma^2} \cdot \frac{1}{1 + \frac{|p_m|^2}{\sigma^2}} \cdot \Re\{p_m^* \Delta_m\}, \nonumber
\end{align}
where the notation $v_m = \max\left(0, R_{\text{th}} - \log_2\left(1 + \frac{|p_m|^2}{\sigma^2}\right)\right)$, $q_p=a^i_{l} c^i_{l} u_i$, $\Delta_m = \mathbf{h}_m^H \mathbf{J}_{i,l} \mathbf{F}_A \mathbf{f}_{D,m}$, and $\mathbf{J}_{i,l}$ is a sparse matrix with its only non-zero element located at the $l$-th entry of the $i$-th column.
Finally, the quadratic coefficient $\mathcal{F}''(0)$ is given by
\begin{align}
\mathcal{F}''(0) &= \frac{1}{2} \sum_{p=1}^{P} \left[ 8(\Re\{z_p^* q_p\})^2 + 4(|z_p|^2 - b_p)|q_p|^2 \right] \nonumber \\
&\quad + \frac{\mu}{2} \sum_{m=1}^{M} \left[ \frac{8(\Re\{p_m^* \Delta_m\})^2}{(\ln 2)^2 \cdot \sigma^4} \cdot \frac{1}{\left(1 + \frac{|p_m|^2}{\sigma^2}\right)^2} \right.   \\
&\quad \left. + \frac{2v_m}{\ln 2} \left( \frac{4(\Re\{p_m^* \Delta_m\})^2}{\left(1 + \frac{|p_m|^2}{\sigma^2}\right)^2 \sigma^4} - \frac{2|\Delta_m|^2}{\left(1 + \frac{|p_m|^2}{\sigma^2}\right) \sigma^2} \right) \right]. \nonumber
\end{align}
The derivation of the quadratic function coefficients is provided in Appendix~\ref{app_3}.
According to the properties of quadratic functions, the optimal radiation amplitude for an element lies within the interval [0,1] only when $0 < -\frac{\mathcal{F}'(0)}{2\mathcal{F}''(0)} < 1$.
Otherwise, the optimal radiation amplitude is attained at the boundary. Owing to the limited range of the feasible region [0,1], the optimal radiation amplitude is typically achieved at the boundary.

\begin{remark}\label{remark3}
In tri-hybrid holographic ISAC, constraining the element radiation amplitude to binary values, i.e., 1-bit amplitude control, serves as a suboptimal solution, thereby significantly reducing both hardware and algorithmic complexity.
\end{remark}

\begin{figure}[t]
\setlength{\abovecaptionskip}{0pt}
\setlength{\belowcaptionskip}{0pt}
	\centering
    \includegraphics[width=0.47\textwidth]{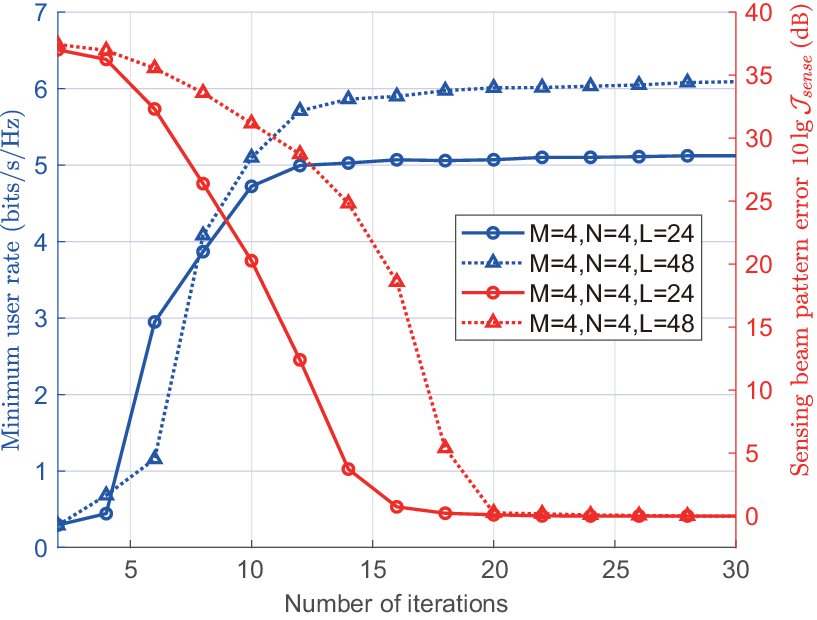}
	\caption{ISAC performance vs. the number of iterations.}
	\label{Fig:iteration}
\vspace{-0em}
\end{figure}

\subsection{Computational Complexity and Convergence}

The digital beamformer $\mathbf{F}_D$ has dimensions of $M \times M$, and the optimization variable $\mathbf{x}$ has a dimension of $2M^2$. The complexity of the SQP algorithm in section~\ref{III-A} primarily arises from gradient computation, with a computational complexity of $\mathcal{O}(PLM + LM^3)$ for gradients associated with $P$ sensing directions and $M$ communication users. In the joint optimization updating $\mathbf{F}_A\in\mathbb{C}^{NM \times M}$ and $\mathbf{F}_E\in\mathbb{C}^{L \times NM}$, the total complexity for updating $\mathbf{F}_E$ is $\mathcal{O}((P + M)LNM)$, while that for updating $\mathbf{F}_A$ is also $\mathcal{O}((P + M)LNM)$.
Therefore, the total computational complexity of the iterative algorithm is $\mathcal{O}((PLM + LM^3)+(P + M)LNM)$.
In the RHS-enabled tri-hybrid beamformer, the computational complexity is predominantly determined by the number of elements $L$, while the values of $M$ and $N$ are relatively small. As a result, the algorithm's complexity scales approximately linearly with $L$, making it well-suited for large-scale RHS-enabled tri-hybrid arrays.

As indicated in~\textbf{\emph{Remark~\ref{remark3}}}, the optimized amplitude responses of the RHS elements typically reside near the boundary values. Adopting 1-bit amplitude control as a suboptimal strategy for the RHS, the amplitude response of 0 or 1 is selected for each element to minimize the objective function.
This can eliminate the need for computing RHS element gradients, thereby reducing the computational complexity of the iterative algorithm.

Fig.~\ref{Fig:iteration} illustrates the number of iterations required for the proposed algorithm to achieve convergence.
For different numbers of antenna elements, both communication and sensing performance enhance and gradually saturate as the tri-hybrid beamformer is optimized, demonstrating the convergence of the proposed algorithm. For RHS element counts of 24 and 48, convergence is achieved within fewer than 20 iterations.

\section{Simulation Results}\label{Sec5}


In the parameter setup, the carrier frequency is 30 GHz, and the wavenumber on the RHS surface is set as $|\mathbf{k}_s|=\frac{2\sqrt{3}\pi}{\lambda}$.
Considering the practical array structure of the RHS, the antenna element spacings are $d_x = \frac{\lambda}{2}$ and $d_y = \frac{\lambda}{4}$.
The azimuth angles for communication users and sensing targets span the range $[0^\circ, 360^\circ]$, while the elevation angles are in $[0^\circ, 90^\circ]$.
In the iterative algorithm, the multiplicative factor is set to 1.5, and the penalty coefficient interval $[\mu_{\min},\mu_{\max}]$ is [1, 1000].
The system is configured with 4 users and 5 sensing targets. The BS employs $M = 4$ RF chains, each connected to only $N = 4$ phase shifters, with no dedicated RF chains allocated for sensing tasks.
The following schemes are considered in the simulation.
\begin{itemize}
\item \textbf{RHS-enabled trihybrid beamforming}: The number of RF chains is $M = 4$, and the number of phase shifters is $M \times N = 16$. The tri-hybrid beamformer optimization follows the proposed scheme in Section~\ref{Sec3}.
\item \textbf{RHS-enabled trihybrid beamforming~(1-bit amplitude control)}: The number of RF chains is $M = 4$, and the number of phase shifters is $M \times N = 16$. Considering the practical hardware constraints of the RHS, the EM-layer RHS elements employ 1-bit amplitude control~(AC).
\item \textbf{RHS-enabled hybrid beamforming}: The signals from the RF chains are directly fed into the RHS array. The number of RF chains is $M = 4$, and the number of RHS elements $L$ is set to be identical to that in the tri-hybrid beamforming scheme.
\item \textbf{PA-enabled hybrid beamforming}: The digital and analog beamformers employ a partially-connected architecture, with $M = 4$ RF chains and a total of $14\times14$ or $18\times18$ antenna elements.
\end{itemize}

\begin{figure}[t!]
    \centering
    
    \subfloat[$R_{th}$ = 4 bit/s/Hz]{
        \includegraphics[width=0.43\textwidth]{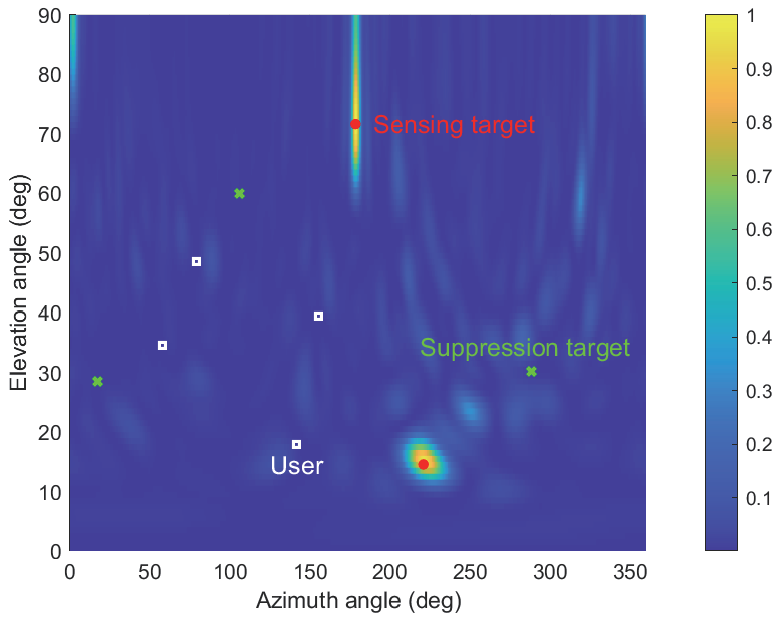}
        \label{fig:sub1}
    }
    
    \vspace{-0.1cm}
    
    \subfloat[$R_{th}$ = 6 bit/s/Hz]{
        \includegraphics[width=0.43\textwidth]{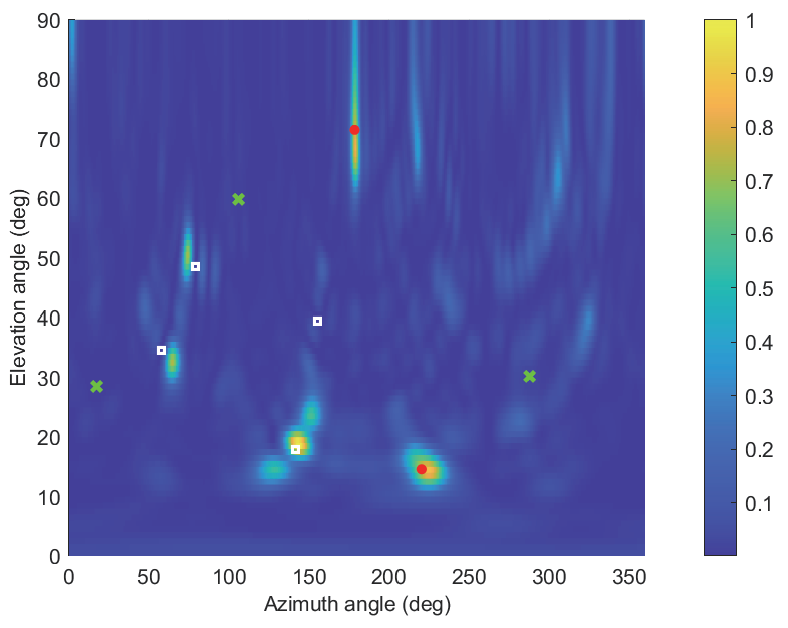}
        \label{fig:sub2}
    }
    
    \vspace{-0.1cm}
    
    \subfloat[$R_{th}$ = 8 bit/s/Hz]{
        \includegraphics[width=0.43\textwidth]{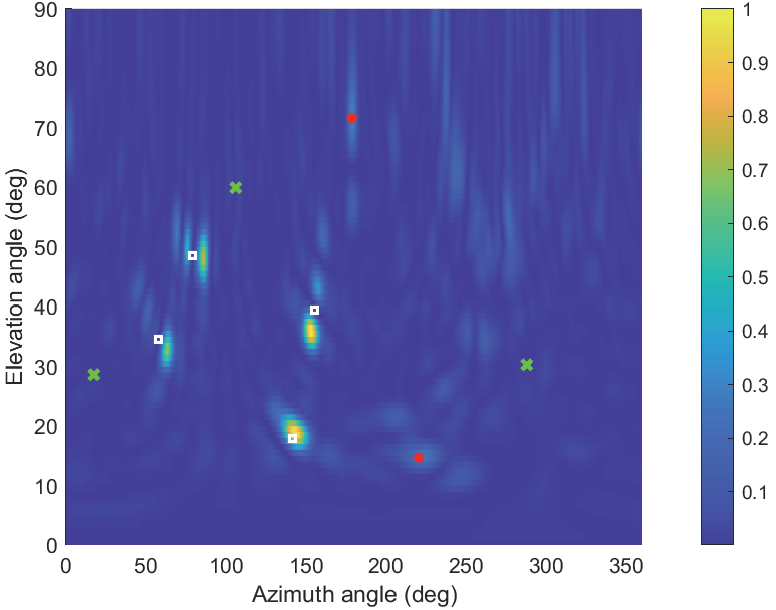}
        \label{fig:sub3}
    }
    
    \caption{White squares indicate user locations, red dots represent sensing targets, and green crosses denote interference suppression targets.
As the threshold rate $R_{th}$ increases, the beam energy is progressively shifted from sensing targets to communication users, reflecting the inherent trade-off between communication and sensing performance.}
    \label{fig:three_figures}
\end{figure}

\subsection{Holographic Patterns of RHS-Enabled Tri-Hybrid ISAC}

Fig.~\ref{fig:three_figures} illustrates the beam patterns of the tri-hybrid holographic ISAC under different threshold rate requirements, with the number of RHS elements $L$ set to 48. In the figure, the white squares indicate user locations, red dots represent sensing detection targets intended for high beamforming gain, and green crosses denote sensing suppression targets where minimal beam gain is desired.
As shown in the figure, the proposed scheme accomplishes the integrated design of communication and sensing beams using limited communication resources, enabling simultaneous beam coverage for both communication users and sensing targets. Moreover, as the threshold rate $R_{th}$ increases, the beam energy gradually shifts from sensing targets to users, indicating a transition from sensing tasks to communication tasks. This demonstrates that the tri-hybrid holographic ISAC achieves a controllable trade-off between communication and sensing performance.

\subsection{Comparison with Traditional ISAC Schemes}

\begin{figure}[t]
\setlength{\abovecaptionskip}{0pt}
\setlength{\belowcaptionskip}{0pt}
	\centering
    \includegraphics[width=0.48\textwidth]{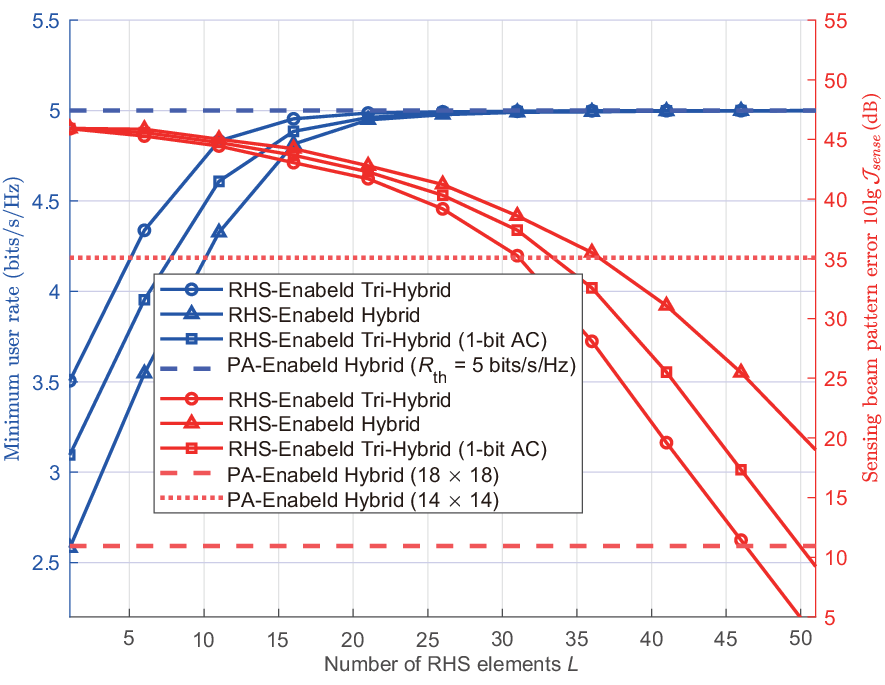}
	\caption{ISAC performance of different schemes.}
	\label{Fig:compare}
\vspace{-0em}
\end{figure}

Under identical transmit power, threshold rate requirements, and sensing target gain constraints, Fig.~\ref{Fig:compare} compares the performance of different ISAC schemes.
The hybrid beamformer utilizes 196 and 324 phase shifters in the analog layer, whereas the tri-hybrid beamformer employs only 16.
As the number of RHS elements $L$ increases, the signal processing capability of the EM-layer beamformer is enhanced, thereby improving both holographic communication and sensing performance. With the expansion of the EM-layer scale, the performance of the proposed tri-hybrid ISAC gradually approaches and surpasses that of the conventional hybrid beamformer, while significantly reducing the number of phase shifters. Compared to the RHS-enabled hybrid beamformer, the proposed scheme achieves superior ISAC performance due to the introduction of subarray-level phase control in the analog layer, as discussed in \textbf{\emph{Remark~\ref{remark2}}}. When 1-bit amplitude control is employed in the EM layer, both hardware and algorithmic complexity are reduced, and the performance approaches that of continuous amplitude control, validating~\textbf{\emph{Remark~\ref{remark3}}}.

\subsection{Effect of Number of RHS elements on Holographic ISAC}

\begin{figure}[t]
\setlength{\abovecaptionskip}{0pt}
\setlength{\belowcaptionskip}{0pt}
	\centering
    \includegraphics[width=0.49\textwidth]{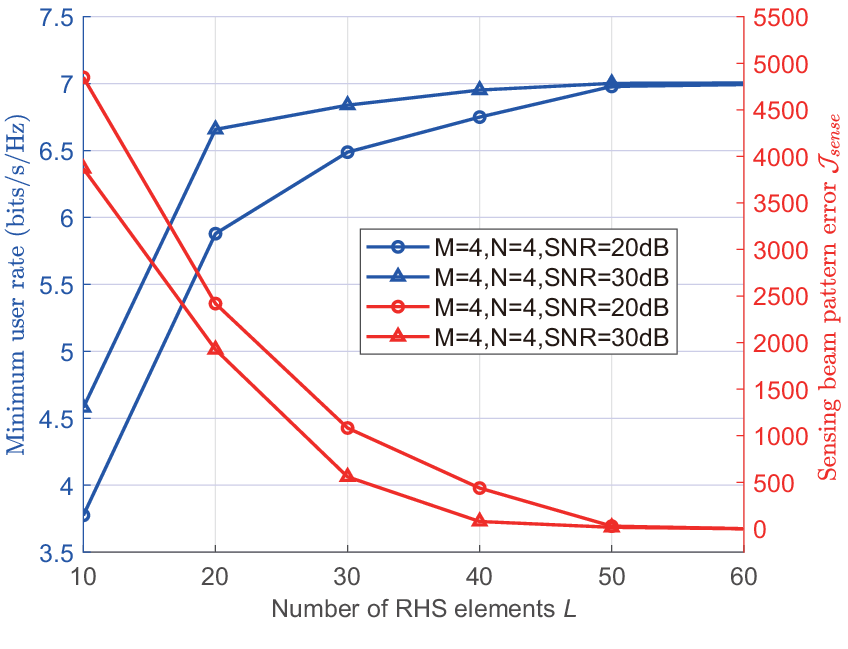}
	\caption{ISAC performance vs. the number of RHS elements.}
	\label{Fig:element}
\vspace{-0em}
\end{figure}

Fig.~\ref{Fig:element} illustrates the variation of the ISAC performance of the tri-hybrid holographic system with respect to the number of RHS elements. For the same number of elements, a higher SNR yields better ISAC performance due to improved communication channel conditions. As the scale of the RHS array expands, its signal processing capability is enhanced, leading to improvements in both communication and sensing performance. However, the rate of performance gain gradually diminishes with the increase in the number of elements. This occurs because the RHS, functioning as a leaky-wave antenna with serial transmission, progressively radiates energy into free space as EM waves propagate along the metasurface. The energy radiated by newly added elements decreases gradually, resulting in a slower improvement in holographic ISAC performance. Moreover, as presented in~\textbf{\emph{Remark~\ref{remark1}}}, once a certain number of elements is reached, the incremental beamforming gain approaches zero, and the ISAC performance tends to saturate.

\subsection{Impact of Subarray-Level Phase Control on Holographic Beamforming}

\begin{figure}[t]
\setlength{\abovecaptionskip}{0pt}
\setlength{\belowcaptionskip}{0pt}
	\centering
    \includegraphics[width=0.48\textwidth]{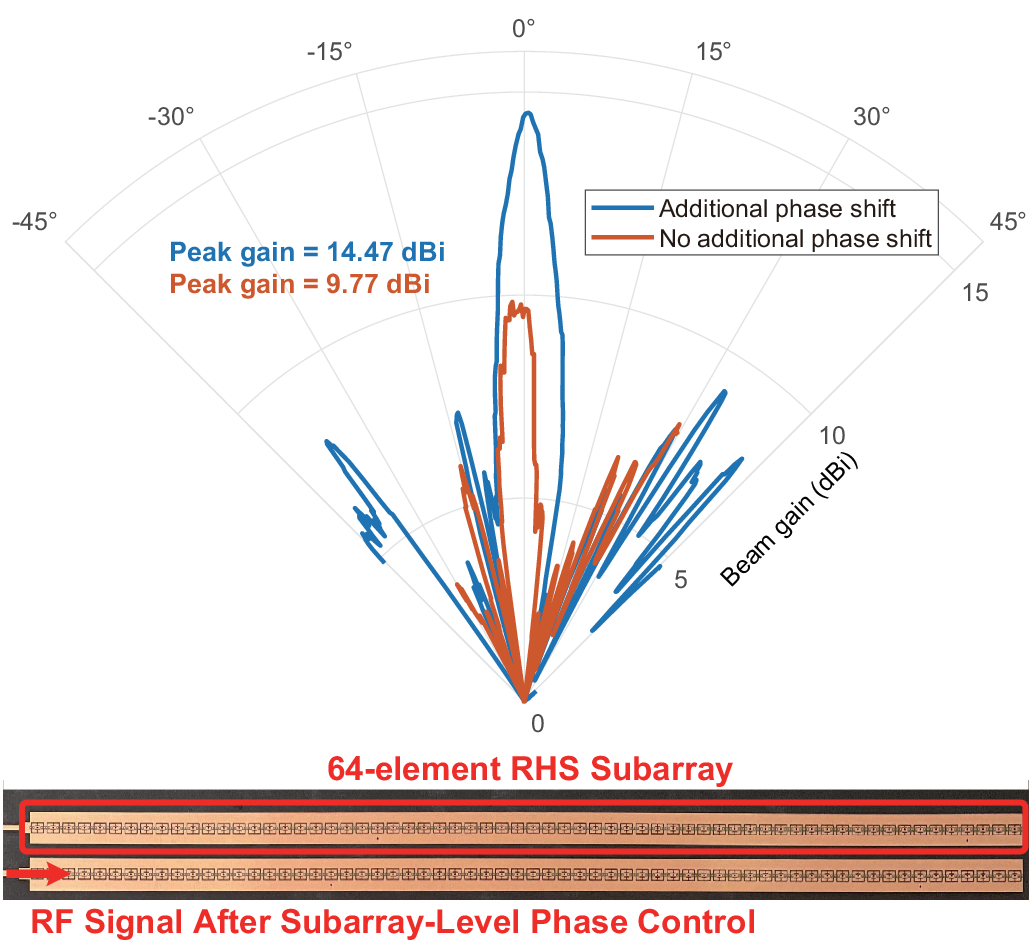}
	\caption{Measured holographic beam gain.}
	\label{Fig:phase}
\vspace{-0em}
\end{figure}

To validate the impact of the introduced subarray-level phase control on holographic beamforming gain, we measured the beam pattern using the RHS prototype.
RHS prototype has dimensions of \(20.25 \times 10.91 \times 0.12  \text{cm}^3\), which consists of 8 subarrays with 64 elements each. The feed is placed on one side of each subarray to inject the reference wave into the RHS.
Fig.~\ref{Fig:phase} shows the measured holographic beam patterns with and without subarray-level phase shifting. To achieve in-phase superposition of the holographic beams from all subarrays and enhance the beamforming gain, the optimized phase shifts are applied to the RF signals before feeding them into the respective subarrays. As illustrated, a holographic beam directed at $0^\circ$ is designed, and the application of subarray-level phase shifts increases the peak beam gain by 4.7 dB. The reference wave feeding the RHS exhibits a fixed phase shift upon reaching each element, while its radiation amplitude can be controlled per element. Introducing subarray-level phase control adds a new degree of freedom, allowing the phase of the EM wave across elements to be adjusted rather than fixed, thereby enabling in-phase superposition of the holographic beams from all subarrays to improve the beamforming gain as~\textbf{\emph{Remark~\ref{remark2}}}.

\section{Conclusions}\label{Sec6}
This paper proposes an RHS-enabled tri-hybrid holographic ISAC scheme, which achieves simultaneous multi-user communication and multi-target sensing aided by large-scale arrays at a low cost. Specifically, a low-complexity tri-hybrid iterative optimization algorithm is designed to solve the sensing waveform design problem under minimum user rate constraints. An adaptive projected gradient method is employed to jointly control the subarray-level phase in the analog layer and the amplitude response of the RHS in the EM layer for holographic ISAC. Moreover, theoretical analysis reveals the improvement of holographic beamforming gain using only a small number of phase shifters. The distribution of the optimized RHS amplitude responses is also theoretically analyzed, based on which a suboptimal 1-bit amplitude control scheme is proposed to further reduce both algorithmic and hardware complexity.

Simulation results demonstrate that: 1) The proposed tri-hybrid holographic ISAC simultaneously enables both multi-user communication and multi-target sensing tasks, which achieves a controllable trade-off between their performance; 2) The subarray-level phase control introduced by employing a small number of phase shifters enhances holographic beamforming gain, and the performance gradually saturates as the number of RHS elements increases; 3) As the number of RHS elements grows, the performance of the tri-hybrid holographic ISAC progressively approaches and surpasses that of conventional hybrid beamforming, while requiring substantially fewer phase shifters.

\appendices
\section{Gradient of $\mathbf{F}_D$}\label{app_1}
\textbf{Sensing term}: The gradient of the $G_p$ term is expressed as
\begin{gather}
\nabla_{\mathbf{F}_D^*} G_p = \left( \nabla_{\mathbf{w}_{\text{sense}}^*} G_p \right) \cdot \left( \nabla_{\mathbf{F}_D^*} \mathbf{w}_{\text{sense}}^* \right),
\end{gather}
where $\nabla_{\mathbf{w}_{\text{sense}}^*} G_p = \mathbf{a}_p \mathbf{a}_p^{H} \mathbf{w}_{\text{sense}}$ and $\mathbf{w}_{\text{sense}}^* = \mathbf{G}^* (\mathbf{F}_D^* \mathbf{1}_M).$
Differentiating with respect to \( \mathbf{w}_{\text{sense}}^* \) as
\begin{equation}
d\mathbf{w}_{\text{sense}}^* = \mathbf{G}^* (d\mathbf{F}_D^* \mathbf{1}_M).
\end{equation}
To find the gradient, we express the differential in terms of the matrix trace. For any matrix
 \( \mathbf{A} \) with the same dimensions as \( d\mathbf{F}_D^* \), we have
\begin{equation}
\text{Tr}(\mathbf{A}^{H} d\mathbf{w}_{\text{sense}}^*) = \text{Tr}(\mathbf{A}^{H} \mathbf{G}^* d\mathbf{F}_D^* \mathbf{1}_M).
\end{equation}
Utilizing the cyclic permutation property of the trace:
\begin{equation}
\text{Tr}(\mathbf{A}^{H} \mathbf{G}^* d\mathbf{F}_D^* \mathbf{1}_M) = \text{Tr}(\mathbf{1}_M \mathbf{A}^{H} \mathbf{G}^* d\mathbf{F}_D^*).
\end{equation}
According to the definition of the Wirtinger derivative, the gradient \( \nabla_{\mathbf{F}_D^*} \mathbf{w}_{\text{sense}}^* \) should satisfy
\begin{equation}
d\mathbf{w}_{\text{sense}}^* = \text{Tr}\left( (\nabla_{\mathbf{F}_D^*} \mathbf{w}_{\text{sense}}^*)^{H} d\mathbf{F}_D^* \right).
\end{equation}
Comparing the two trace expressions, we obtain:
\begin{equation}
\nabla_{\mathbf{F}_D^*} \mathbf{w}_{\text{sense}}^* = (\mathbf{1}_M \mathbf{A}^{H} \mathbf{G}^*)^{\mathsf{T}} = \mathbf{G}^{{H}} \mathbf{A} \mathbf{1}_M^{{T}}.
\end{equation}
Since $\mathbf{A}$ is an arbitrary test matrix, it is replaced by the preceding term \( \nabla_{\mathbf{w}_{\text{sense}}^*} G_p \) in the chain rule.
We obtain the final gradient expression:
\begin{equation}
\frac{\partial G_p}{\partial \mathbf{F}_D^*} =  \mathbf{F}_A^H \mathbf{F}_E^H \mathbf{a}_p \mathbf{a}_p^H \mathbf{F}_E\mathbf{F}_A \sum_{m=1}^M\mathbf{f}_{D,m} \mathbf{1}_M^T.
\end{equation}

\textbf{Communication term}:
$\text{SINR}_k$ is defined as $\frac{S_k}{I_k}$, and the power term and the interference term are expressed as
\begin{align}
S_k &= |\mathbf{u}_k^{{H}} \mathbf{f}_{D,k}|^2 = \mathbf{f}_{D,k}^{{H}} \mathbf{u}_k \mathbf{u}_k^{{H}} \mathbf{f}_{D,k}, \\
I_k &= \sum_{m \neq k} |\mathbf{u}_k^{{H}} \mathbf{f}_{D,m}|^2 + \sigma^2 = \sum_{m \neq k} \mathbf{f}_{D,m}^{{H}} \mathbf{u}_k \mathbf{u}_k^{{H}} \mathbf{f}_{D,m} + \sigma^2,
\end{align}
where $\mathbf{u}_k = \mathbf{F}_A^H \mathbf{F}_E^H \mathbf{h}_k$.
According to the quotient rule, the derivative of $\text{SINR}_k$ with respect to \( \mathbf{f}_{D,j}^* \) is given by
\begin{equation}
\frac{\partial \text{SINR}_k}{\partial \mathbf{f}_{D,j}^*} = \frac{ \frac{\partial S_k}{\partial \mathbf{f}_{D,j}^*} \cdot I_k - S_k \cdot \frac{\partial I_k}{\partial \mathbf{f}_{D,j}^*} }{ I_k^2 }.
\end{equation}
\textbf{Case 1: \( j = k \)}. The term $\frac{\partial S_k}{\partial \mathbf{f}_{D,k}^*} = \mathbf{u}_k \mathbf{u}_k^{{H}} \mathbf{f}_{D,k}$, and $\frac{\partial I_k}{\partial \mathbf{f}_{D,k}^*} = \mathbf{0}$.
\begin{align}
\frac{\partial \text{SINR}_k}{\partial \mathbf{f}_{D,k}^*} &= \frac{ (\mathbf{u}_k \mathbf{u}_k^{{H}} \mathbf{f}_{D,k}) \cdot I_k - S_k \cdot \mathbf{0} }{ I_k^2 } \\
&= \frac{ \mathbf{u}_k \mathbf{u}_k^{{H}} \mathbf{f}_{D,k} }{ I_k }.
\end{align}
\textbf{Case 2: \( j \neq k \)}.
The term $\frac{\partial S_k}{\partial \mathbf{f}_{D,j}^*} = \mathbf{0}$.
The interference is expressed as
\begin{align}
I_k &= \sum_{m \neq k} \mathbf{f}_{D,m}^{{H}} \mathbf{u}_k \mathbf{u}_k^{{H}} \mathbf{f}_{D,m} + \sigma^2, \\
\frac{\partial I_k}{\partial \mathbf{f}_{D,j}^*} &= \frac{\partial}{\partial \mathbf{f}_{D,j}^*} \left( \mathbf{f}_{D,j}^{{H}} \mathbf{u}_k \mathbf{u}_k^{{H}} \mathbf{f}_{D,j} \right)  \\
&= \mathbf{u}_k \mathbf{u}_k^{{H}} \mathbf{f}_{D,j}. 
\end{align}
Hence, the derivative is given by
\begin{align}
\frac{\partial \text{SINR}_k}{\partial \mathbf{f}_{D,j}^*} &= \frac{ \mathbf{0} \cdot I_k - S_k \cdot (\mathbf{u}_k \mathbf{u}_k^{{H}} \mathbf{f}_{D,j}) }{ I_k^2 } \\
&= -\frac{S_k}{I_k^2} \mathbf{u}_k \mathbf{u}_k^{{H}} \mathbf{f}_{D,j}.
\end{align}

\section{Gradient of $\mathbf{F}_A \text{ and } \mathbf{F}_E$}\label{app_2}
\textbf{Sensing term}:
Based on the quotient rule, the derivative of ${\mathcal{J}_{\text{sense}}}$ with respect to $\mathbf{F}_A^*$ is given by
\begin{equation}
\frac{\partial \mathcal{J}_{\text{sense}}}{\partial \mathbf{F}_A^*} = \sum_{p=1}^{P} 2\delta_p \cdot \frac{\partial G_p}{\partial \mathbf{F}_A^*},
\end{equation}
where $G_p = \mathbf{f}_{D}^H \mathbf{G}^H \mathbf{a}_p \mathbf{a}_p^H \mathbf{G} \mathbf{f}_{D}$. Its differential is expressed as
\begin{equation}\label{app_eq1}
dG_p = \mathbf{f}_{D}^H d(\mathbf{G}^H) \mathbf{a}_p \mathbf{a}_p^H \mathbf{G} \mathbf{f}_{D} + \mathbf{f}_{D}^H \mathbf{G}^H \mathbf{a}_p \mathbf{a}_p^H d(\mathbf{G}) \mathbf{f}_{D}.
\end{equation}
Substituting $d\mathbf{G} = \mathbf{F}_{E} \cdot d\mathbf{F}_A$ and $d\mathbf{G}^H = d\mathbf{F}_A^H \cdot \mathbf{F}_{E}^H$ into \eqref{app_eq1} yields
\begin{gather}
dG_p = \mathbf{f}_{D}^H d\mathbf{F}_A^H \mathbf{F}_{E}^H \mathbf{a}_p \mathbf{a}_p^H \mathbf{G} \mathbf{f}_{D},\\
\frac{\partial G_p}{\partial \mathbf{F}_A^*} = \mathbf{F}_{E}^H \mathbf{a}_p \mathbf{a}_p^H \mathbf{G} \mathbf{f}_{D} \mathbf{f}_{D}^H.
\end{gather}
Taking the real part of the differential with respect to $G_p$ as
\begin{equation}\label{app_eq2}
dG_p = 2\Re\left\{ \mathbf{f}_{D}^H \mathbf{G}^H \mathbf{a}_p \mathbf{a}_p^H d(\mathbf{G}) \mathbf{f}_{D} \right\}.
\end{equation}
Substituting $d\mathbf{G} = (d\mathbf{F}_{E,\text{amp}} \odot \mathbf{C}_{\text{phase}}) \cdot \mathbf{F}_A$ into \eqref{app_eq2} yields
\begin{gather}
dG_p = 2\Re\left\{ \mathbf{f}_{D}^H \mathbf{G}^H \mathbf{a}_p \mathbf{a}_p^H (d\mathbf{F}_{E,\text{amp}} \odot \mathbf{C}_{\text{phase}}) \mathbf{F}_A \mathbf{f}_{D} \right\},\\
\frac{\partial G_p}{\partial \mathbf{F}_{E,\text{amp}}} = 2\Re\left\{ \mathbf{C}_{\text{phase}}^* \odot \left[ \mathbf{a}_p \mathbf{a}_p^H \mathbf{G} \mathbf{f}_{D} \mathbf{f}_{D}^H \mathbf{F}_A^H \right] \right\}.
\end{gather}

\textbf{Communication term}:
According to the quotient rule, the derivative of $\text{SINR}_k$ with respect to $\mathbf{F}_A^*$ is given by
\begin{equation}
\frac{\partial \text{SINR}_k}{\partial \mathbf{F}_A^*} = \frac{1}{I_k} \frac{\partial S_k}{\partial \mathbf{F}_A^*} - \frac{S_k}{I_k^2} \frac{\partial I_k}{\partial \mathbf{F}_A^*}.
\end{equation}
Let \( g_k = \mathbf{h}_k^H \mathbf{G} \mathbf{f}_{D,k} \), whose derivative with respect to \(\mathbf{F}_A^*\) is given by
\begin{equation}
\frac{\partial g_k}{\partial \mathbf{F}_A^*} = 0, \quad \frac{\partial g_k^*}{\partial \mathbf{F}_A^*} = \mathbf{F}_E^H \mathbf{h}_k \mathbf{f}_{D,k}^H.
\end{equation}
Therefore, the derivative of \( S_k \) is expressed as
\begin{equation}
\frac{\partial S_k}{\partial \mathbf{F}_A^*} = \frac{\partial (g_k g_k^*)}{\partial \mathbf{F}_A^*} = g_k \frac{\partial g_k^*}{\partial \mathbf{F}_A^*} = g_k \mathbf{F}_E^H \mathbf{h}_k \mathbf{f}_{D,k}^H.
\end{equation}
The gradient of the interference term is given by
\begin{equation}
\frac{\partial I_k}{\partial \mathbf{F}_A^*} = \sum_{j \neq k} g_{k,j} \mathbf{F}_E^H \mathbf{h}_k \mathbf{f}_{D,j}^H,
\end{equation}
where the term \(g_{k,j} = (\mathbf{h}_k^H \mathbf{G} \mathbf{f}_{D,j})^*\).

Rewriting $G_p$ as $\mathbf{f}_{D}^H \mathbf{G}^H \mathbf{a}_p \mathbf{a}_p^H \mathbf{G} \mathbf{f}_{D}$, its differential is expressed as
\begin{align}
dG_p &= \mathbf{f}_{D}^H d(\mathbf{G}^H) \mathbf{a}_p \mathbf{a}_p^H \mathbf{G} \mathbf{f}_{D} + \mathbf{f}_{D}^H \mathbf{G}^H \mathbf{a}_p \mathbf{a}_p^H d(\mathbf{G}) \mathbf{f}_{D}, \\
&= 2\Re\left\{ \mathbf{f}_{D}^H \mathbf{G}^H \mathbf{a}_p \mathbf{a}_p^H d(\mathbf{G}) \mathbf{f}_{D} \right\}.
\end{align}
Substituting $d\mathbf{G} = (d\mathbf{F}_{E,\text{amp}} \odot \mathbf{C}_{\text{phase}}) \cdot \mathbf{F}_A$ yields
\begin{equation}
\frac{\partial G_p}{\partial \mathbf{F}_{E,\text{amp}}} = 2\Re\left\{ \mathbf{C}_{\text{phase}}^* \odot \left[ \mathbf{a}_p \mathbf{a}_p^H \mathbf{G} \mathbf{f}_{D} \mathbf{f}_{D}^H \mathbf{F}_A^H \right] \right\}.
\end{equation}

Using the quotient rule as
\begin{equation}
\frac{\partial \text{SINR}_k}{\partial \mathbf{F}_{E,\text{amp}}} = \frac{1}{I_k} \frac{\partial S_k}{\partial \mathbf{F}_{E,\text{amp}}} - \frac{S_k}{I_k^2} \frac{\partial I_k}{\partial \mathbf{F}_{E,\text{amp}}},
\end{equation}
where \(S_k\) is a real-valued function. Taking the real part yields
\begin{equation}
\frac{\partial S_k}{\partial \mathbf{F}_{E,\text{amp}}} = 2 \cdot \Re \left\{ \frac{\partial g_k}{\partial \mathbf{F}_{E,\text{amp}}} g_k^* \right\}.
\end{equation}
Expanding \(g_k\) as $\mathbf{h}_k^H (\mathbf{F}_{E,\text{amp}} \odot \mathbf{C}_{\text{phase}} \cdot \mathbf{F}_A) \mathbf{f}_{D,k}$, for an arbitrary element:
\begin{equation}
\frac{\partial g_k}{\partial [\mathbf{F}_{E,\text{amp}}]_{m,n}} = \mathbf{h}_k^H \left( \frac{\partial \mathbf{G}}{\partial [\mathbf{F}_{E,\text{amp}}]_{m,n}} \right) \mathbf{f}_{D,k},
\end{equation}
where the term $\frac{\partial \mathbf{G}}{\partial [\mathbf{F}_{E,\text{amp}}]_{m,n}} = \mathbf{J}_{m,n} \odot \mathbf{C}_{\text{phase}} \cdot \mathbf{F}_A$, and $\mathbf{J}_{m,n}$ is the sparse matrix.
Combining into matrix form yields 
\begin{gather}
\frac{\partial g_k}{\partial \mathbf{F}_{E,\text{amp}}} = \mathbf{C}_{\text{phase}} \odot (\mathbf{h}_k \mathbf{f}_{D,k}^H \mathbf{F}_A^H),\\
\frac{\partial S_k}{\partial \mathbf{F}_{E,\text{amp}}} = 2 \cdot \Re \left\{ \mathbf{C}_{\text{phase}} \odot \left[ \mathbf{h}_k \mathbf{f}_{D,k}^H \mathbf{F}_A^H g_k^* \right] \right\}.
\end{gather}

The gradient of the interference term is given by
\begin{gather}
\frac{\partial I_k}{\partial \mathbf{F}_{E,\text{amp}}} = \sum_{j \neq k}  2 \cdot \Re \left\{ \frac{\partial g_{k,j}}{\partial \mathbf{F}_{E,\text{amp}}} g_{k,j}^* \right\},\\
\frac{\partial g_{k,j}}{\partial \mathbf{F}_{E,\text{amp}}} = \mathbf{C}_{\text{phase}} \odot (\mathbf{h}_k \mathbf{f}_{D,j}^H \mathbf{F}_A^H),\\
\frac{\partial I_k}{\partial \mathbf{F}_{E,\text{amp}}} = \sum_{j \neq k} 2 \cdot \Re \left\{ \mathbf{C}_{\text{phase}} \odot \left[ \mathbf{h}_k \mathbf{f}_{D,j}^H \mathbf{F}_A^H g_{k,j}^* \right] \right\}.
\end{gather}

\section{Coefficients of the Quadratic Function}\label{app_3}

\textbf{Linear Coefficient}: The linear coefficient corresponds to the gradient of the objective function at $a=0$:
\begin{equation}
\mathcal{F}'(0) = \frac{\partial \mathcal{J}_{\text{sense}}}{\partial a_l^i}(0) + \mu \frac{\partial \mathcal{V}_{\text{rate}}}{\partial a_l^i}(0),
\end{equation}
where the sensing component is given by
\begin{align}
\frac{\partial \mathcal{J}_{\text{sense}}}{\partial a} &= \sum_{p=1}^{P} 2(G_p - b_p) \frac{\partial G_p}{\partial a_l^i} \\
\frac{\partial G_p}{\partial a_l^i} &= 2\Re\left\{ (z_p + a_l^i q_p)^* q_p \right\} \\
\Rightarrow \frac{\partial \mathcal{J}_{\text{sense}}}{\partial a_l^i}(0) &= \sum_{p=1}^{P} 4(|z_p|^2 - b_p) \Re\{z_p^* q_p\}.
\end{align}
The communication component is expressed as
\begin{align}
\frac{\partial \mathcal{V}_{\text{rate}}}{\partial a^i_{l}} &= \sum_{m=1}^{M} \frac{\partial}{\partial a^i_{l}} \left[ \max\left(0, R_{\text{th}} - R_m\right)^2 \right] \\
&= \sum_{m=1}^{M} -2 v_m \frac{\partial R_m}{\partial a^i_{l}} \\
\frac{\partial R_m}{\partial a^i_{l}} &= \frac{1}{\ln 2} \cdot \frac{1}{1 + \frac{P_m}{\sigma^2}} \cdot \frac{1}{\sigma^2} \cdot \frac{\partial P_m}{\partial a^i_{l}} \\
P_m(a^i_{l}) &= |p_m + a^i_{l} \Delta_m|^2 \\
\frac{\partial P_m}{\partial a^i_{l}} &= 2\Re\{ (p_m + a^i_{l} \Delta_m)^* \Delta_m \} \\
\Rightarrow \frac{\partial R_m}{\partial a^i_{l}}(0) &= \frac{2}{\ln 2 \cdot \sigma^2} \cdot \frac{1}{1 + \frac{|p_m|^2}{\sigma^2}} \cdot \Re\{p_m^* \Delta_m\}.
\end{align}
Therefore, the total linear coefficient is
\begin{align}
\mathcal{F}'(0) &= \sum_{p=1}^{P} 4(|z_p|^2 - b_p) \Re\{z_p^* q_p\} \\
&- \mu \sum_{m=1}^{M} \frac{4 v_m}{\ln 2 \cdot \sigma^2} \cdot \frac{1}{1 + \frac{|p_m|^2}{\sigma^2}} \cdot \Re\{p_m^* \Delta_m\}. \nonumber
\end{align}

\textbf{Quadratic Coefficient}:
The quadratic coefficient corresponds to half of the second derivative of the objective function evaluated at $a^i_{l}=0$:
\begin{equation}
\frac{1}{2} \mathcal{F}''(0) = \frac{1}{2} \left[ \frac{\partial^2 \mathcal{J}_{\text{sense}}}{\partial (a^i_{l})^2}(0) + \mu \frac{\partial^2 \mathcal{V}_{\text{rate}}}{\partial (a^i_{l})^2}(0) \right].
\end{equation}
For simplicity, the superscripts and subscripts of $a^i_{l}$ are omitted.
The sensing component is given by
\begin{align}
\frac{\partial^2 \mathcal{J}_{\text{sense}}}{\partial a^2} &= \sum_{p=1}^{P} \left[ 2\left(\frac{\partial G_p}{\partial a}\right)^2 + 2(G_p - b_p) \frac{\partial^2 G_p}{\partial a^2} \right] \nonumber \\
\frac{\partial G_p}{\partial a}(0) &= 2\Re\{z_p^* q_p\}, \frac{\partial^2 G_p}{\partial a^2} = 2|q_p|^2 \\
\Rightarrow \frac{\partial^2 \mathcal{J}_{\text{sense}}}{\partial a^2}(0) &= \sum_{p=1}^{P} \left[ 8(\Re\{z_p^* q_p\})^2 + 4(|z_p|^2 - b_p)|q_p|^2 \right] \nonumber. 
\end{align}
The communication component is given by
\begin{align}
\frac{\partial^2 \mathcal{V}_{\text{rate}}}{\partial a^2} &= \sum_{m=1}^{M} \left[ 2\left(\frac{\partial R_m}{\partial a}\right)^2 - 2v_m \frac{\partial^2 R_m}{\partial a^2} \right]\\
\frac{\partial R_m}{\partial a}(0) &= \frac{2}{\ln 2 \cdot \sigma^2} \cdot \frac{1}{1 + \frac{|p_m|^2}{\sigma^2}} \cdot \Re\{p_m^* \Delta_m\} \\
\frac{\partial^2 R_m}{\partial a^2} &= \frac{1}{\ln 2} \left[\frac{1}{\sigma^2(1 + \frac{P_m}{\sigma^2})} \frac{\partial^2 P_m}{\partial a^2} -\frac{\left(\frac{\partial P_m}{\partial a}\right)^2}{\sigma^4\left(1 + \frac{P_m}{\sigma^2}\right)^2} \right] \nonumber\\
\frac{\partial P_m}{\partial a}(0) &= 2\Re\{p_m^* \Delta_m\}, \frac{\partial^2 P_m}{\partial a^2} = 2|\Delta_m|^2 \\
\Rightarrow \frac{\partial^2 R_m}{\partial a^2}(0) &= \frac{1}{\ln 2} \left[ -\frac{4(\Re\{p_m^* \Delta_m\})^2}{\left(1 + \frac{|p_m|^2}{\sigma^2}\right)^2 \sigma^4} + \frac{2|\Delta_m|^2}{\left(1 + \frac{|p_m|^2}{\sigma^2}\right) \sigma^2} \right]. \nonumber
\end{align}
Therefore, the total quadratic coefficient is
\begin{align}
\mathcal{F}''(0) &= \frac{1}{2} \sum_{p=1}^{P} \left[ 8(\Re\{z_p^* q_p\})^2 + 4(|z_p|^2 - b_p)|q_p|^2 \right] \nonumber \\
&\quad + \frac{\mu}{2} \sum_{m=1}^{M} \left[ \frac{8(\Re\{p_m^* \Delta_m\})^2}{(\ln 2)^2 \cdot \sigma^4} \cdot \frac{1}{\left(1 + \frac{|p_m|^2}{\sigma^2}\right)^2} \right.   \\
&\quad \left. + \frac{2v_m}{\ln 2} \left( \frac{4(\Re\{p_m^* \Delta_m\})^2}{\left(1 + \frac{|p_m|^2}{\sigma^2}\right)^2 \sigma^4} - \frac{2|\Delta_m|^2}{\left(1 + \frac{|p_m|^2}{\sigma^2}\right) \sigma^2} \right) \right]. \nonumber
\end{align}

\end{document}